\documentclass[aps,superscriptaddress,amsmath,amssymb,floatfix,twocolumn,showpacs,amsfonts,longbibliography]{revtex4-2}
\usepackage{natbib}
\setcitestyle{numbers,square}
\usepackage{times}
\usepackage[varg]{txfonts}
\usepackage{textcomp}
\usepackage{graphicx}
\usepackage{subfigure}
\usepackage{color}
\usepackage[colorlinks=true,citecolor=blue,urlcolor=blue,linkcolor=blue,hyperindex]{hyperref}
\usepackage{braket}
\usepackage{overpic}
\usepackage{bm}

\def\non{\nonumber}

\begin{document}

\title{Type-II Dirac points and Dirac nodal loops on the magnons of square-hexagon-octagon lattice}
\author{Meng-Han Zhang}
\affiliation{State Key Laboratory of Optoelectronic Materials and Technologies, Center for Neutron Science and Technology, Guangdong Provincial Key Laboratory of Magnetoelectric Physics and Devices, School of Physics, Sun Yat-Sen University, Guangzhou 510275, China}

\author{Dao-Xin Yao}
\email[Corresponding author:]{yaodaox@mail.sysu.edu.cn}
\affiliation{State Key Laboratory of Optoelectronic Materials and Technologies, Center for Neutron Science and Technology, Guangdong Provincial Key Laboratory of Magnetoelectric Physics and Devices, School of Physics, Sun Yat-Sen University, Guangzhou 510275, China}
\affiliation{International Quantum Academy, Shenzhen 518048, China}
\date{\today}

\begin{abstract}

We study topological magnons on an anisotropic square-hexagon-octagon (SHO) lattice which has been found by a two-dimensional Biphenylene network (BPN). We propose the concepts of type-II Dirac magnonic states where new schemes to achieve topological magnons are unfolded without requiring the Dzyaloshinsky-Moriya interactions (DMIs). In the ferromagnetic states, the topological distinctions at the type-II Dirac points along with one-dimensional (1D) closed lines of Dirac magnon nodes are characterized by the Berry phase and the $\mathbb{Z}_2$ invariant. We find pair annihilation of the Dirac magnons and use the Wilson loop method to depict the topological protection of the band-degeneracy. The Green's function approach is used to calculte chiral edge modes and magnon density of states (DOS). We introduce the DMIs to gap the type-II Dirac magnon points and demonstrate the Dirac nodal loops (DNLs) are robust against the DMIs within a certain parameter range. The topological phase diagram of magnon bands is given via calculating the Berry curvature and Chern number. We find that the anomalous thermal Hall conductivity gives connection to the magnon edge current. Furthermore, we derive the differential gyromagnetic ratio to exhibit the Einstein-de Haas effect (EdH) of magnons with topological features.

\end{abstract}

\maketitle
\section{INTRODUCTION}\label{sec:intro}

Recently, a new network of carbon atoms known as the BPN has been synthesized~\cite{Scienceabg4509}, which has the point group $D_{2h}$ with $4-$, $6-$ and $8-$ membered rings~\cite{nanolett2c00528}, which can be considered as an effective SHO lattice where the hexagonal cluster forms the unit cell and the square-octagon is its connection. It can also be visualized as an interpolating hexagon lattice between the honeycomb lattice and the kagome lattice, where the octagons act as defects in the honeycomb structure. Magnons are charge-neutral bosonic quasiparticles representing the collective excitations of the spin waves. The topological magnetic spin excitations in insulating ordered quantum magnets hold the DMIs as the spin-orbit interactions to break the inversion symmetry ~\cite{JPhysChem1958,PhysRev.120.91}. We construct an effective magnonic Hamiltonian of the SHO lattice in the reciprocal space within the Holstein-Primakoff (HP) representation and the Fourier transformation. Our model has six magnon bands which resembles twice variant bands of the kagome lattice retaining partial features of the flat band. We propose the concept of the type-II Dirac crossing points on the stable flat bands in the context of the SHO lattice. The nontrivial topology is achieved via generating gaps to split the nonlinear relationship between the spin magnetization energy and the incident angle. Our results are expected to be experimentally detected in the magnetic materials with the SHO lattice by inelastic neutron scattering, Raman spectroscopy, resonant inelastic X-ray scattering, etc~\cite{PhysRevB.3.157}.

The Einstein-de Haas effect is a phenomenon involving the transfer of angular momentum between microscopic magnetic moments and macroscopic mechanical rotation~\cite{1915EdH}. It provides an accurate measurement of the gyromagnetic ratio than electron-spin resonance or ferromagnetic resonance~\cite{PhysRev.6.239,PhysRevB.79.104410}, which revealed that the origin of magnetism was the intrinsic angular momentum of electrons~\cite{PhysRevLett.112.085503,PhysRevB.99.064428}. Currently, the EdH effect attracts increasing attention and has important applications in the fields of ultrafast magnetism~\cite{RevModPhys.90.015005,PhysRevLett.118.117203} and nano-magneto-mechanical systems~\cite{PhysRevLett.94.167201,PhysRevLett.104.027202,PhysRevB.95.134447,PhysRevB.75.014430}. Meanwhile, the Berry curvature gives rise to anomalous transport phenomenon refering to the generation of a transverse thermal Hall conductivity~\cite{RevModPhys.82.3045}. In a system with non-trivial topology, the temperature gradient applied along one direction induces a transverse heat current flowing in a perpendicular direction~\cite{PhysRevLett171133}. This anomalous thermal Hall effect conductivity $\kappa_{xy}$ is closely connected to the Berry curvature and the heat currents are carried by charge-neutral quasiparticles such as magnons~\cite{PhysRevLett.104.066403,PhysRevLett.120.097702}.

In this work, we show that magnonic SHO lattice analogous to 2D BPN have type-II Dirac states~\cite{PhysRevB.95.075133} and Dirac nodal loops. The type-II Dirac magnonic state can exhibit novel phenomena such as flat bands and topological insulators, allowing for new potential applications. Our proposed scheme is based on a SHO lattice model with only nearest-neighbour exchange couplings, which can realize topologically nontrivial transition by varying the exchange coupling parameters. The topological characteristics at the degenerate band structure are revealed by the $\mathbb{Z}_2$ invariant. We use real-space Green's function approach to calculate of the chiral edge modes and magnon DOS. By introducing the DMIs to gap the energy bands, the magnon Chern number can be defined which reveals the topological features of these magnon modes. We analyse the topological phase diagram of Chern numbers and realize the DNLs which are robust within a certain parameter range of the DMIs.  The topological transport properties are considered via calculating magnon thermal Hall conductivity, where the thermal fluctuations are described by the Curie temperature. We derive the differential gyromagnetic ratio response with the effective-mass tensor to exhibit the EdH effect of magnons. Further studies are needed to explore the potential applications of these novel topological features and gain deeper understandings of the physics behind the EdH effect.

\section{MODEL AND METHODS}\label{sec:model}
\subsection{Spin Model}\label{subsec:SM}

\begin{figure}[t]
\centering
{
\subfigure[]{
\includegraphics[width=1.6 in]{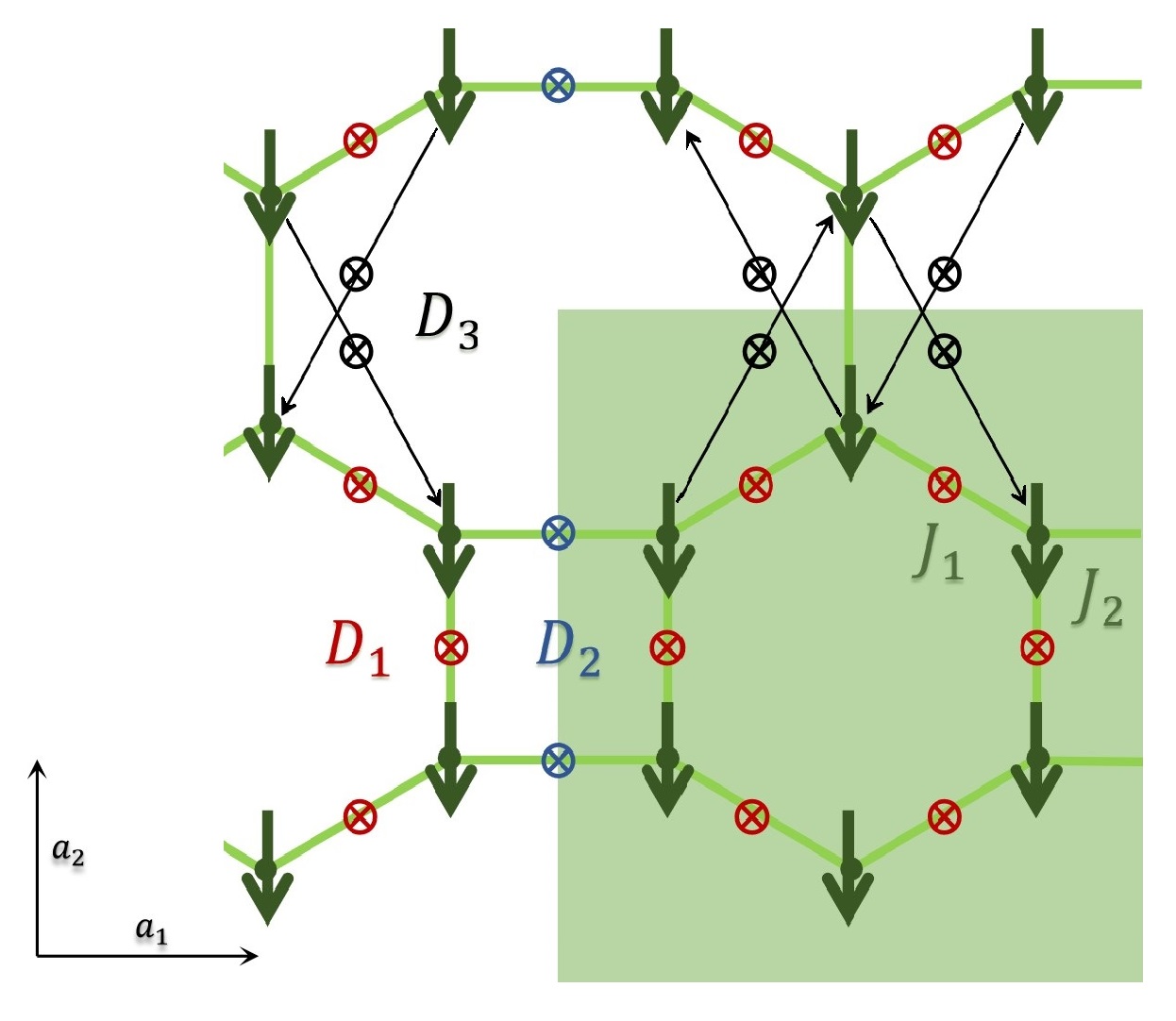}\label{fig1a}}
}
{
\subfigure[]{
\includegraphics[width=1.6 in]{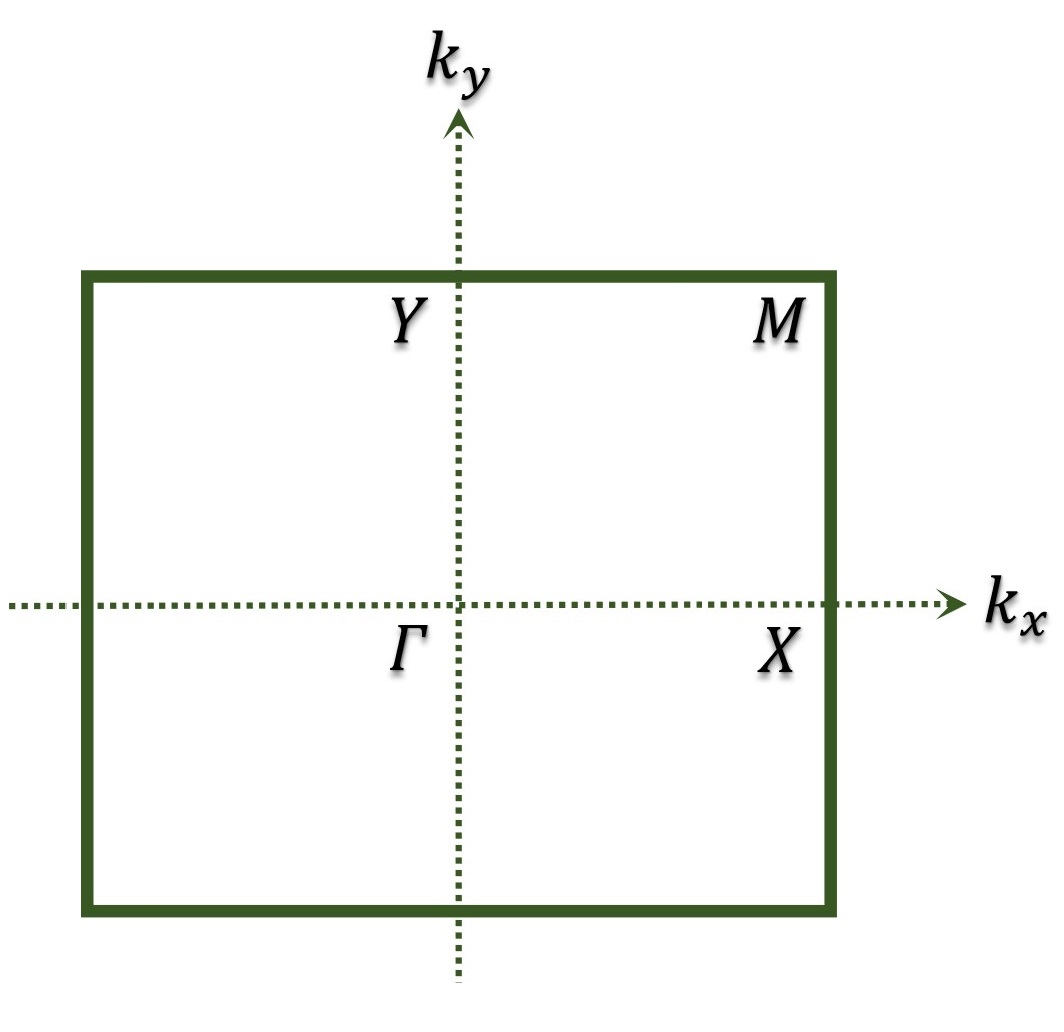}\label{fig1b}}
}
\caption{(a) Schematics of the SHO lattice with shaded region that represent the unit cell. (b)The anisotropic Brillouin zone of the SHO lattice with $2$/$3$ in length and $2$/$(1+\sqrt{3})$ in width. The DMIs are allowed at the midpoints of the nearest neighbours (NN) and the next nearest neighbours (NNN) indicated by crossed circles which leads to fictitious magnetic flux in momentum space.}
\label{Fig1}
\end{figure}

We consider a Heisenberg model on the SHO lattice with six spins in the unit cell as shown in Fig.~\ref{Fig1}, where the total Hamiltonian is given by
\begin{align}
\mathcal{H}=\mathcal{H}_0 +\mathcal{H}_{\mathrm{DM}} +\mathcal{H}_{K} +\mathcal{H}_B,
\end{align}
Our model Hamiltonian contains the nearest neighbour Heisenberg exchange interactions, where the $\mathcal{H}_0$ is
\begin{align}
\mathcal{H}_0=-J_1\sum_{\langle mn\rangle} \boldsymbol{S}_m \cdot \boldsymbol{S}_n -J_2\sum_{\langle mn\rangle} \boldsymbol{S}_m \cdot \boldsymbol{S}_n,
\end{align}
and $J_1$, $J_2$ are two types of the nearest-neighbor exchange couplings within the hexagonal unit cells and between them as shown in Fig.~\ref{Fig1}. The $\mathcal{H}_{\mathrm{DM}}$ terms include two types of nearest-neighbor DMIs and a next nearest-neighbor DMIs in the octagon sublattice. Therefore, it can be considered as
\begin{align}
\mathcal{H}_\mathrm{DM}^{}&= \sum_{\langle mn\rangle_1} \boldsymbol{D}_{1}^{} \cdot (\boldsymbol{S}_m^{} \times \boldsymbol{S}_n^{}) + \sum_{\langle mn\rangle_2} \boldsymbol{D}_{2}^{} \cdot (\boldsymbol{S}_m^{} \times \boldsymbol{S}_n^{}) \non\\
&+\sum_{\langle \langle mn\rangle \rangle} \boldsymbol{D}_{3}^{} \cdot (\boldsymbol{S}_m^{} \times \boldsymbol{S}_n^{}).
\end{align}

\begin{figure*}[t]
\centering
{
\subfigure[]{
\includegraphics[width=1.65 in]{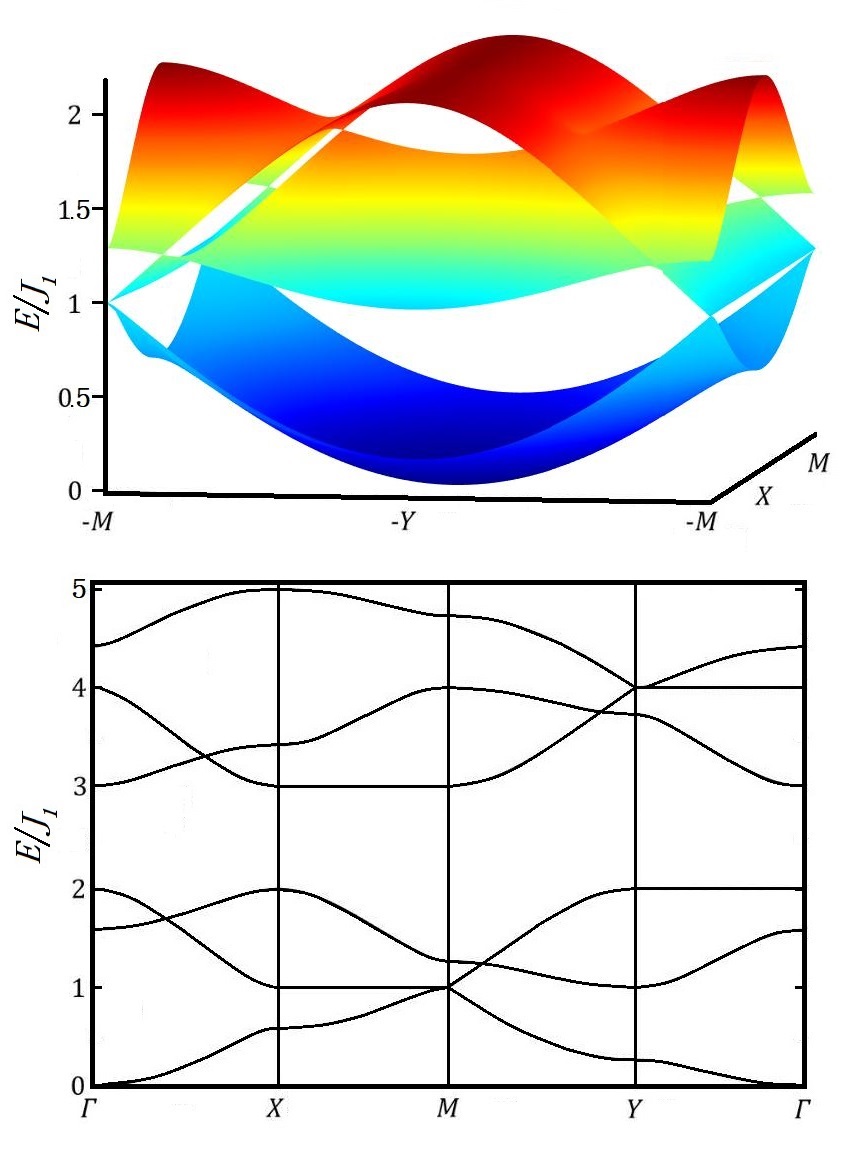}\label{fig2a}}
}
{
\subfigure[]{
\includegraphics[width=1.6 in]{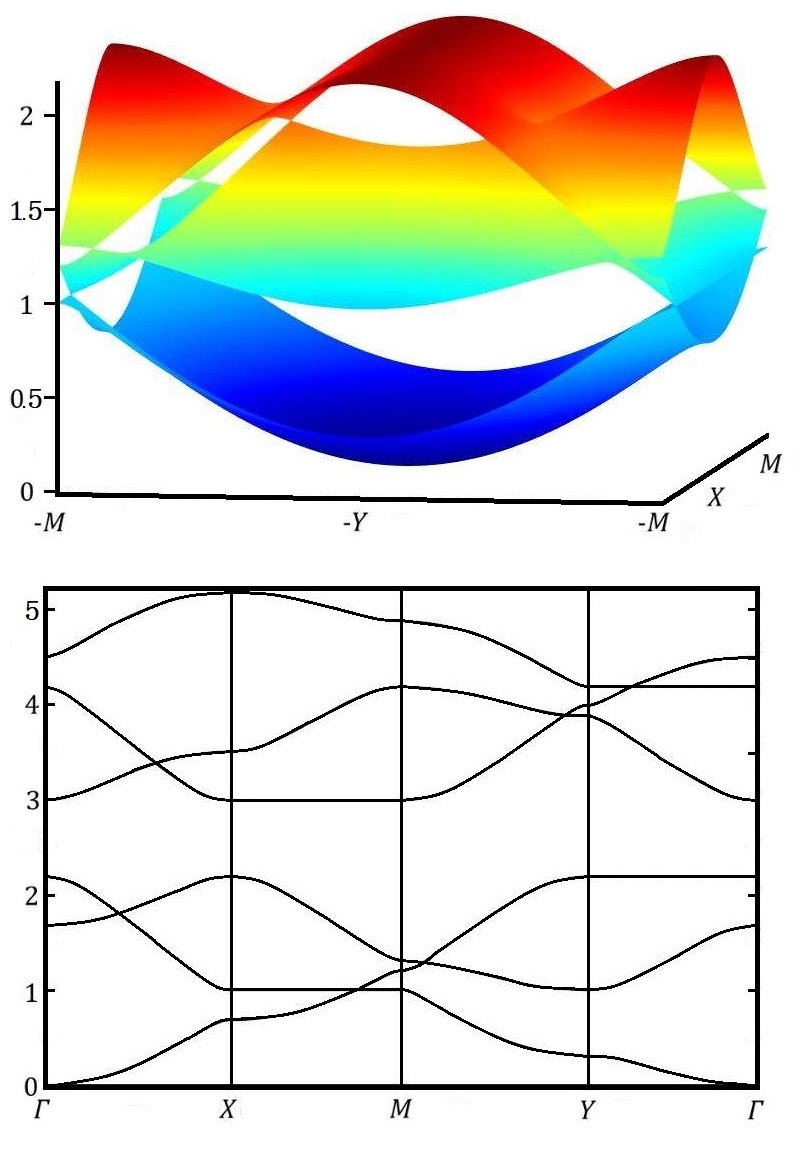}\label{fig2b}}
}
{
\subfigure[]{
\includegraphics[width=1.6 in]{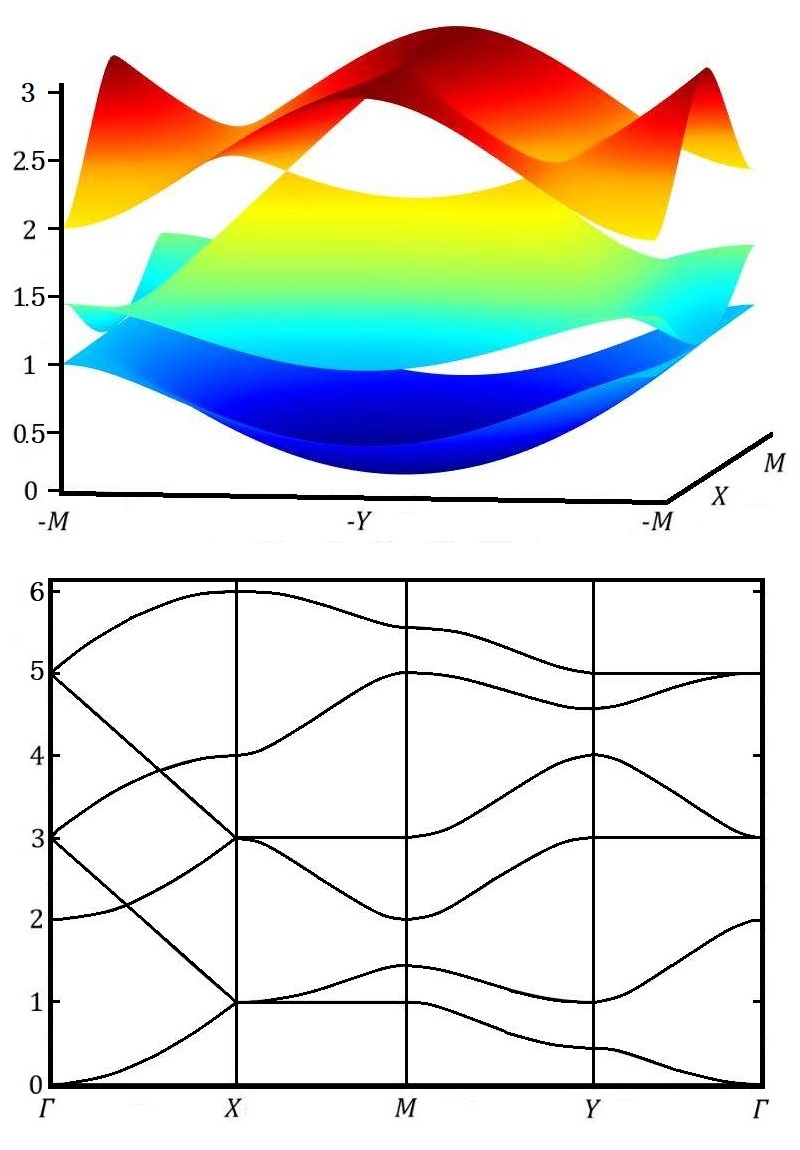}\label{fig2c}}
}
{
\subfigure[]{
\includegraphics[width=1.6 in]{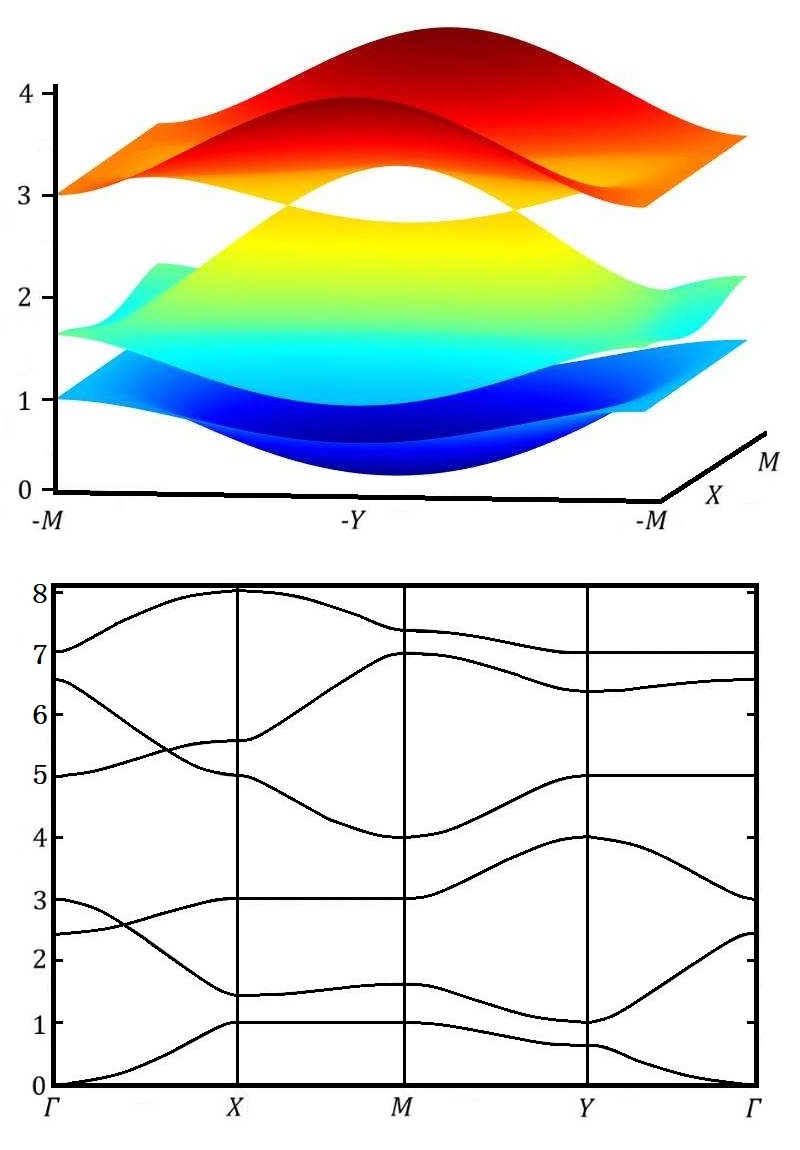}\label{fig2d}}
}
\caption{The magnon energy bands of the SHO lattice and the type-II Dirac magnonic points with various $J_{2}$ interactions. The parameters of ferromagnetic ground state with $D_{1}$=$D_{2}$=$D_{3}$=$K$=$h$=0 are set as: (a)$J_{1}$=1 and $J_{2}$=0.5; (b)$J_{1}$=1 and $J_{2}$=0.6; (c)$J_{1}$=$J_{2}$=1; (d)$J_{1}$=1 and $J_{2}$=2 .}
\label{Fig2}
\end{figure*}

We also introduce the anisotropic term and the Zeeman term to keep the magnetic order even at finite temperature based on the Mermin-Wagner theorem~\cite{EPL47004,PhysRevLett171133}. The anisotropic term is given by
\begin{align}
\mathcal{H}_{K}=-K\sum_{\langle m\rangle} (S_m^z)^2,
\end{align}
where $K$ is the anisotropy along the $z$-axis. And the external Zeeman magnetic field term is given by
\begin{align}
\mathcal{H}_B=-h\sum_{\langle m\rangle} S_m^z,
\end{align}
where $h =g \mu_B B$, $B$ is the external magnetic field.
The SHO has a ferromagnetic ground state for $J_{1}$$>$0 and $J_{2}$$>$0. We use the Holstein-Primakoff (HP) representation to study the magnetic excitations for the ordered states. The original spin Hamiltonian can be mapped to a bosonic model following the HP transformation:
\begin{align}
S^+_m&=S^x_m+iS^y_m=\sqrt{2S-\alpha^{\dag}_m\alpha_m}\alpha^{}_m,\non\\
S^-_m&=S^x_m-iS^y_m=\alpha^{\dag}_m\sqrt{2S-\alpha^{\dag}_m\alpha_m},\non\\
S^z_m&=S-\alpha^{\dag}_m\alpha^{}_m.
\end{align}
where $\alpha^{\dag}_m(\alpha_m^{})$ is the bosonic magnon creation (annihilation) operator at site $m$. Within the approximation of $\sqrt{2S-\alpha^{\dag}_m\alpha_m}$ $\rightarrow$$\sqrt{2S}$, the Hamiltonian has the form
\begin{align}
\mathcal{H}&=-\Big[\sum_{\langle mn\rangle_1}(J_1+i\nu_{mn}D_1)S\alpha^{\dag}_m\alpha^{}_n+\sum_{\langle mn\rangle_2}(J_2+i\nu_{mn}D_2)S\alpha^{\dag}_m\alpha^{}_n\non\\
&+\sum_{\langle \langle mn\rangle\rangle}(i\nu_{mn}D_3)S\alpha^{\dag}_m\alpha^{}_n +H.c.\Big] +(2K+h)\sum_{\langle m\rangle}\alpha^{\dag}_m\alpha^{}_m+E^{}_0,
\end{align}
where $E_0$ is ground state energy. And the $\nu_{mn}=\pm1$ depends on the direction of DMIs. Subsequently, we perform the Fourier transformation using the definition
\begin{align}
\alpha^{\dag}_{\boldsymbol{k}}=\frac{1}{\sqrt{N}}\sum_me^{i\boldsymbol{k}\cdot\boldsymbol{R}_m}\alpha^{\dag}_m.
\end{align}
Thus, in the reciprocal space the Hamiltonian is given by
\begin{align}
\mathcal{H}=\sum_{\boldsymbol{k}}\psi^{\dag}_{\boldsymbol{k}}H(\boldsymbol{k})\psi^{}_{\boldsymbol{k}},
\end{align}
where $\psi^{\dag}_{\boldsymbol{k}}=(\alpha^{\dag}_{1,\boldsymbol{k}},\alpha^{\dag}_{2,\boldsymbol{k}},\alpha^{\dag}_{3,\boldsymbol{k}},\alpha^{\dag}_{4,\boldsymbol{k}},\alpha^{\dag}_{5,\boldsymbol{k}},\alpha^{\dag}_{6,\boldsymbol{k}})$. The spin wave Hamiltonian matrix is
\begin{align}
S\left[
\begin{array}{ccc}
A_{\boldsymbol{k}} & B_{\boldsymbol{k}} \\
B_{\boldsymbol{k}}^{\dag} & A_{\boldsymbol{-k}}^{\dag} \\
\end{array}
\right],
\end{align}
with matrix $A_{\boldsymbol{k}}$ is
\begin{align}
\left[
\begin{array}{ccc}
E & -\gamma_3e^{i\boldsymbol{k\cdot(\sqrt{3}a_1+3a_2)}} & -\gamma_3e^{-i\boldsymbol{k\cdot(\sqrt{3}a_1-3a_2)}} \\
-\gamma_3e^{-i\boldsymbol{k\cdot(\sqrt{3}a_1+3a_2)}} & E & -\gamma_2e^{i\boldsymbol{k\cdot 2a_1}} \\
-\gamma_3e^{i\boldsymbol{k\cdot(\sqrt{3}a_1-3a_2)}} & -\gamma_2e^{-i\boldsymbol{k\cdot 2a_1}} & E \\
\end{array}
\right],
\end{align}
and matrix $B_{\boldsymbol{k}}$ is
\begin{align}
\left[
\begin{array}{ccc}
-J_2e^{-i\boldsymbol{k\cdot 2a_2}} & -\gamma_1e^{-i\boldsymbol{k\cdot(\sqrt{3}a_1+a_2)}} & -\gamma_1e^{i\boldsymbol{k\cdot(\sqrt{3}a_1-a_2)}} \\
-\gamma_1e^{-i\boldsymbol{k\cdot(\sqrt{3}a_1+a_2)}} & 0 & -\gamma_1e^{i\boldsymbol{k\cdot 2a_2}} \\
-\gamma_1e^{i\boldsymbol{k\cdot(\sqrt{3}a_1-a_2)}} & -\gamma_1e^{i\boldsymbol{k\cdot 2a_2}} & 0 \\
\end{array}
\right],
\end{align}
where $E=2J_1+J_2+2K+h$, $\gamma_1=J_1+i\nu_{mn}D_1$, $\gamma_2=J_2+i\nu_{mn}D_2$ and $\gamma_3=i\nu_{mn}D_3$. The lattice vectors are given by $\boldsymbol{a_1}=\frac{1}{2}(1, 0)a$ and $\boldsymbol{a_2}=\frac{1}{2}(0, 1)a$ with the lattice constant chosen as $a$=0.1nm.

\subsection{Green's Functions and Magnon Density of States}\label{subsec:CNTH}

\begin{figure}[t]
\centering
{
\subfigure[]{
\includegraphics[width=3.4 in]{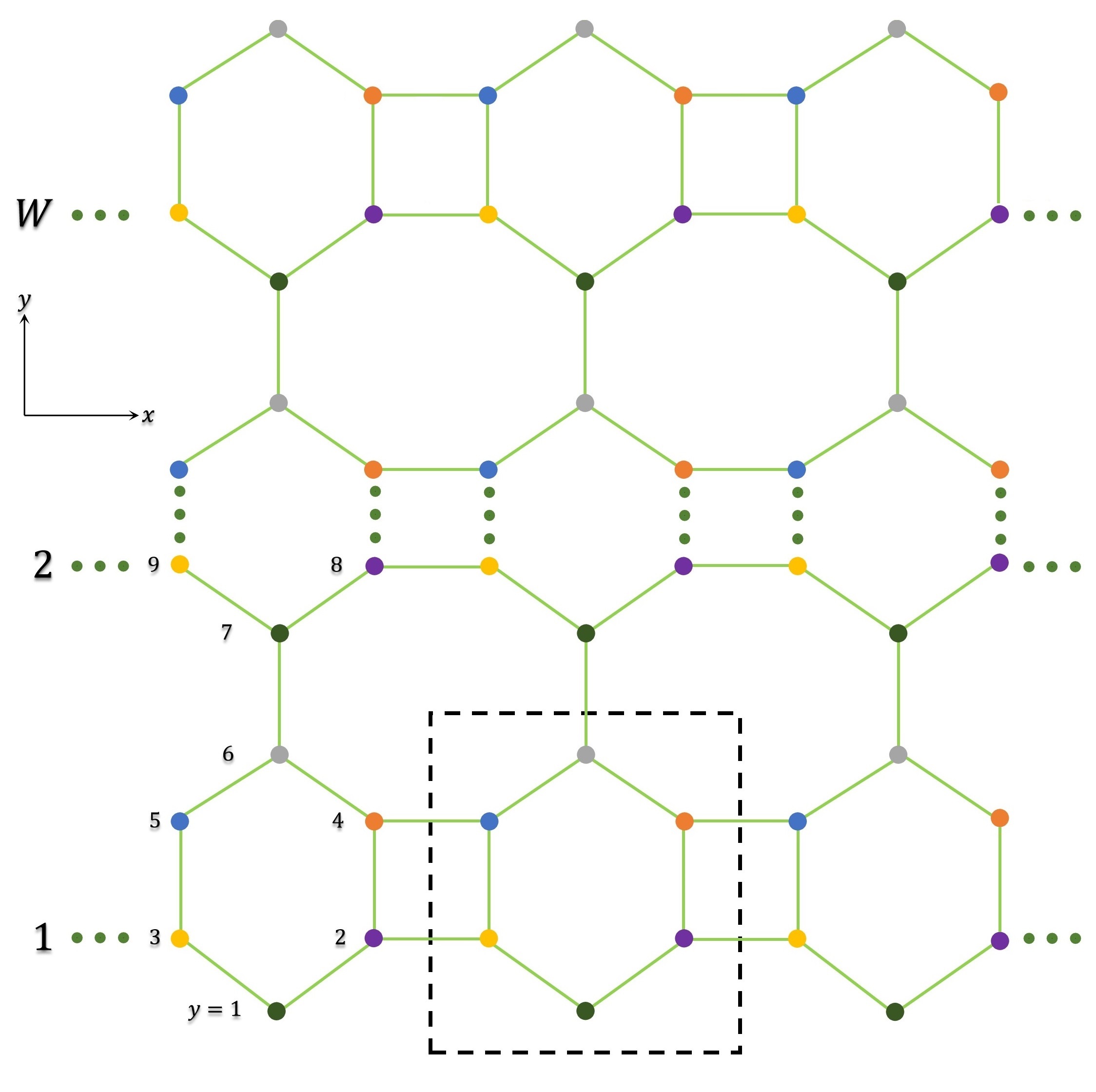}\label{fig3a}}
}
{
\subfigure[]{
\includegraphics[width=3.4 in]{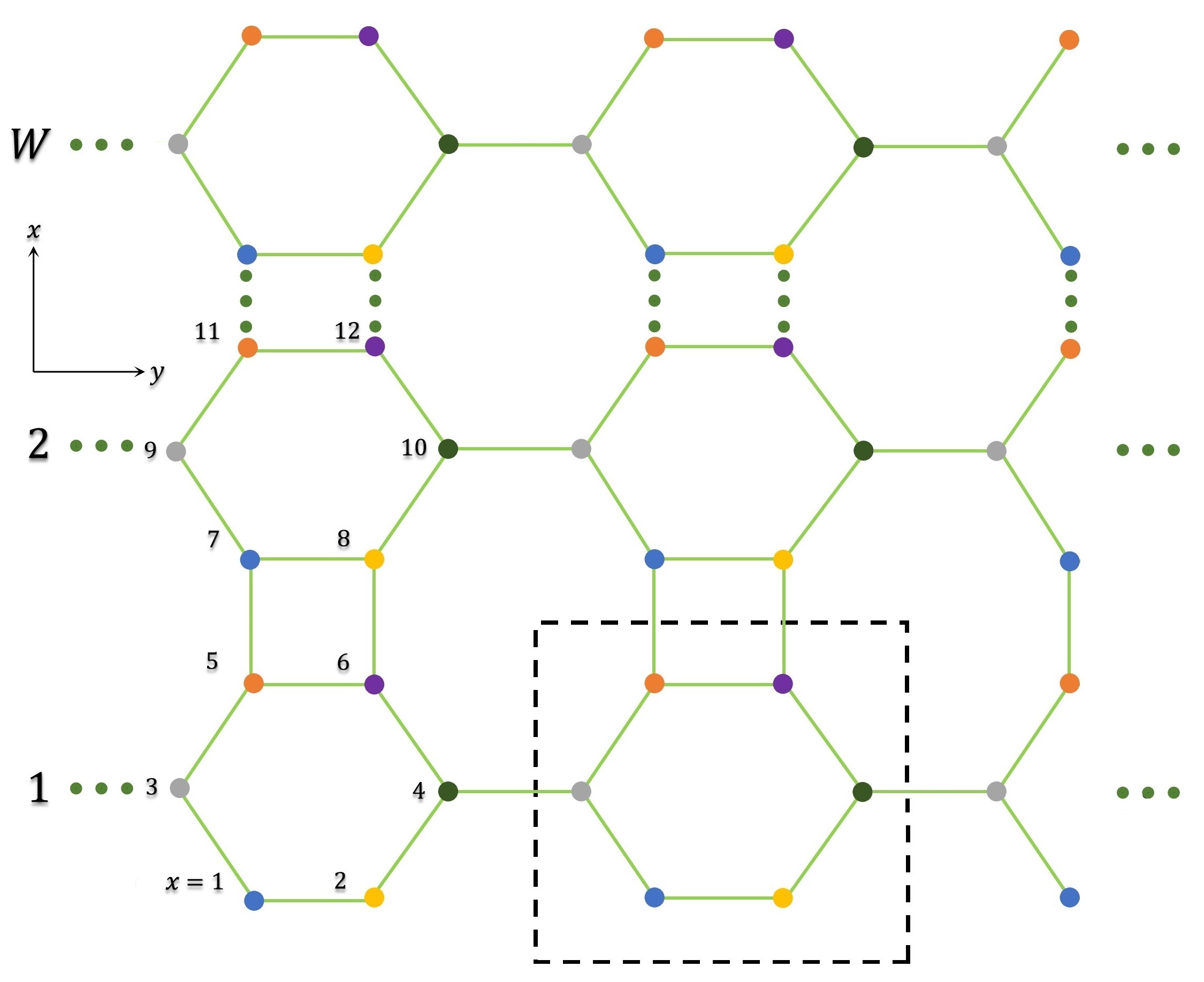}\label{fig3b}}
}
\caption{The SHO ribbon has W periodic one-dimensional chains. (a) The numbers nearing sites are $y$ indices with periodic boundary condition along $x$-axis and open boundary condition along $y$-axis; (b) The numbers nearing sites are $x$ indices with periodic boundary condition along $y$-axis and open boundary condition along $x$-axis.}
\label{Fig3}
\end{figure}

We rewrite our Hamiltonian in $(k_x, y)$ space along the $y$ direction as shown in Fig.~\ref{fig3a}
\begin{align}
\alpha^{\dag}_{ky}=\frac{1}{\sqrt{N_{x}}}\sum_{m}e^{ik\boldsymbol{R}_{m}\cdot\boldsymbol{e}_{x}}\alpha^{\dag}_{my}.
\end{align}

According to the anisotropic edge modes, we also give the Hamiltonian for $(x, k_y)$ space (see Fig.~\ref{fig3b})
\begin{align}
\alpha^{\dag}_{kx}=\frac{1}{\sqrt{N_{y}}}\sum_{m}e^{ik\boldsymbol{R}_{m}\cdot\boldsymbol{e}_{y}}\alpha^{\dag}_{mx},
\end{align}
where $y$ or $x$ runs from $i_1$ to 6($W$-1)+$i_1$ ($i_1$=$\{$1, 2, 3, 4, 5, 6,$\}$) and $W$ denotes the number of periodic 1D chains. We replace $k_x$ or $k_y$ by $k$. The formalism for calculating the band structure of the ribbon geometry is a $6W\times6W$ matrix-form Hamiltonian which is given by
\begin{align}
\mathcal{H}=\sum_{k}\varphi^{\dag}_{k}H(k)\varphi^{}_{k},
\end{align}
where $\varphi^{\dag}_{k}=(\alpha^{\dag}_{i_{1},k}, \alpha^{\dag}_{i_{1}+1,k},..., \alpha^{\dag}_{6(W-1)+i_{1},k})$ in the open boundary condition $\alpha^{\dag}_{0,k}|$0$\rangle$=$\alpha^{\dag}_{6W+1,k}|$0$\rangle$=0. The Hamiltonian matrix can be written as
\begin{align}
H(k)=
\left[
\begin{array}{ccccc}
G(k) & F(k)^{\dag} & 0 & \cdots & 0\\
F(k) & G(k) & F(k)^{\dag} & \ddots & \vdots \\
0 & F(k) & \ddots & \ddots & 0 \\
\vdots & \ddots & \ddots & \ddots & F(k)^{\dag} \\
0 & \cdots & 0 & F(k) & G(k) \\
\end{array}
\right],
\end{align}
where $G(k)$ and $F(k)$ are 6$\times$6 matrices with $G(k)_{ii}$=$E$ (i=$\{$1, 2, 3, 4, 5, 6$\}$), $G(k)_{ij}$=$G(k)^{\dag}_{ji}$, $G(k)_{15}$ = $G(k)_{16}$ = $G(k)_{24}$ = $G(k)_{26}$ = $G(k)_{34}$ = $G(k)_{35}$ = $-\gamma_1$. As shown in Fig.~\ref{fig3a}, $G(k)_{23}$ = $-\gamma_2e^{ika_3}$, $G(k)_{56}$ = $-\gamma_2e^{-ika_3}$, $F(k)_{12}$ = $F(k)_{46}$ = $-\gamma_3e^{\frac{\sqrt{3}}{2}ika_3}$, $F(k)_{13}$ = $F(k)_{45}$ = $-\gamma_3e^{-\frac{\sqrt{3}}{2}ika_3}$, $F(k)_{14}$ = $-J_2$, $G(k)_{ij}$ = 0 (otherwise), $F(k)_{ij}$ = 0 (otherwise), $a_3$ = 0.5$a$. The top and bottom edges are perpendicular to the $y$ direction. As shown in Fig.~\ref{fig3b}, $G(k)_{12}$ = $G(k)_{13}$ = $-\gamma_3e^{\frac{3}{2}ika_3}$, $G(k)_{14}$ = $-J_2e^{ika_3}$, $G(k)_{45}$ = $G(k)_{46}$ = $-\gamma_3e^{-\frac{3}{2}ika_3}$, $F(k)_{23}$ = $F(k)_{56}$ = $-\gamma_2$, $G(k)_{ij}$ = 0 (otherwise), $F(k)_{ij}$ = 0 (otherwise), $a_3$ = 0.5$a$. The top and bottom edges are perpendicular to the $x$ direction. We choose $W$=$30$ to ensure that the results are convergent with $W$.

For the purpose of calculating transport properties of magnons, we introduce the retarded and advanced Green's functions.
\begin{align}
G^R(r, r')=\sum_{k,n} \frac{\alpha^{\dag}_{k,n}(r')\alpha^{}_{k,n}(r)}{\varepsilon+i\eta-H},G^A(r, r')=[G^R(r, r')]^{\dag},
\end{align}
where $\eta$ is a positive infinitesimal, $\varepsilon$ is the excitation energy, $r$ and $r'$ represent excitation and response respectively.
The spectral representation of the Green's function can be written as~\cite{PhysRevB.90.024412}
\begin{align}
A=\sum_{k,n} \alpha^{}_{k,n}(r)\alpha^{\dag}_{k,n}(r')\frac{2\eta}{(\varepsilon-H)^2+\eta^2}.
\end{align}

And the magnon DOS can also be defined as
\begin{align}
\mathbf{\rho}(\varepsilon) = \sum_{k,n} \alpha^{}_{k,n} \alpha^{\dag}_{k,n} \delta(\varepsilon-H) =\frac{\hbar \textmd{Tr}(A)}{2 \pi}.
\end{align}

\subsection{Wilson loop and Kubo Formula}\label{subsec:CNTH}

In the type-II Dirac points, the Berry phase can distinguish between trivial and nontrivial topological phases associated with the band crossings. We acquire the Berry phase over a loop by integrating the Berry connection of band structures.

\begin{align}
A_{n}^{\lambda}=i\langle\psi_{\lambda}|\nabla_{\boldsymbol{k}_{n}}|\psi_{\lambda}\rangle.
\end{align}
with $|\psi_{\lambda}\rangle$ being the normalized wave function of the $\lambda$th Bloch band such that $H(\boldsymbol{k})|\psi_{\lambda}\rangle=E_{\lambda}(\boldsymbol{k})|\psi_{\lambda}\rangle$.

Using the Wilson loop method associated with the Berry's connection, the Berry phase is defined for a closed path encircling the type-II Dirac point nodes in the Brillouin zone (BZ) with $\gamma$=$\pi$, whereas $\gamma$=0 otherwise.

\begin{align}
\gamma_{\boldsymbol{n}}^{\lambda}= \oint_{\partial BZ} A_{n}^{\lambda} \cdot d\boldsymbol{k}.
\end{align}

By manipulating the parameters of the DMIs, nontrivial band topology can be characterized by a nonzero Berry curvature and generally topological invariant like Chern number. The form of Berry curvature is given by

\begin{align}
\Omega_{\lambda \boldsymbol{k}}=i \sum_{n \neq \lambda}\frac{[{\langle\ \lambda\left| {\nabla_{\boldsymbol{k}} H(\boldsymbol{k})}\right| n}\rangle \times {\langle\ n \left| {\nabla_{\boldsymbol{k}} H(\boldsymbol{k})}\right| \lambda}\rangle]_z }{(E_\lambda-E_n)^2},
\end{align}

The associated Chern number assigned to the $n$th band is defined by

\begin{align}
C_{n} = \frac{1}{2\pi} \int_{BZ} d^2 k \Omega_{n \boldsymbol{k}}.
\end{align}

The Chern number denotes the topological nature of reciprocal space with a gap in the magnon spectrum, which protects against scattering and other perturbations.

\subsection{Curie temperature and Thermal Hall Conductivity}\label{subsec:CTHC}
The thermal fluctuations can cause the sublattice magnetization to deviate from its saturation value, which can be described by the Curie temperature $T_c$. Taking into account the magnetization along the $z$-axis, the deviation of the SHO lattice is defined as
\begin{align}
\Delta m=S-\langle S^z_m\rangle = \langle\alpha^{\dag}_m\alpha_m\rangle=\sum_{n,\boldsymbol{k}} \rho(\varepsilon_{n \boldsymbol{k}}),
\end{align}
where the $T_c$ is determined by $\Delta m$ ($T_c$)=$S$. As the temperature approaches the $T_c$, thermal energy increases and causes increased random motion of the magnetic moments for the SHO lattice. Under the $T_c$, the intrinsic anomalous thermal Hall conductivity can be written as $\kappa_{xy}$ with a weighted summation of the Berry curvature~\cite{EPL47004,RevModPhys.82.3045}.
\begin{align}
\kappa_{xy}=-\frac{k_B^2 T}{4\pi^2 \hbar a} \sum_{n,\boldsymbol{k}} c_2[\rho(\varepsilon_{n \boldsymbol{k}})] \Omega_{n \boldsymbol{k}},
\end{align}
where $k_B$ is the Boltzmann constant, $T$ is the temperature and $\rho(\varepsilon_{n \boldsymbol{k}})$ $=[e^{\varepsilon_{n \boldsymbol{k}}/k_{B}T}-1]^{-1}$ is the Bose function. We choose the lattice constant $a$=0.1nm as the typical layer spacing for practical calculation. The $c_{2}(x)$ is given by
\begin{align}
c_2=(1+x)(\ln\frac{1+x}{x})^2-(\ln x)^2-2Li_2(-x),
\end{align}
where $Li_2(x)$ is the polylogarithmic function.

\begin{figure}[t]
\centering
{
\subfigure[]{
\includegraphics[width=1.6 in]{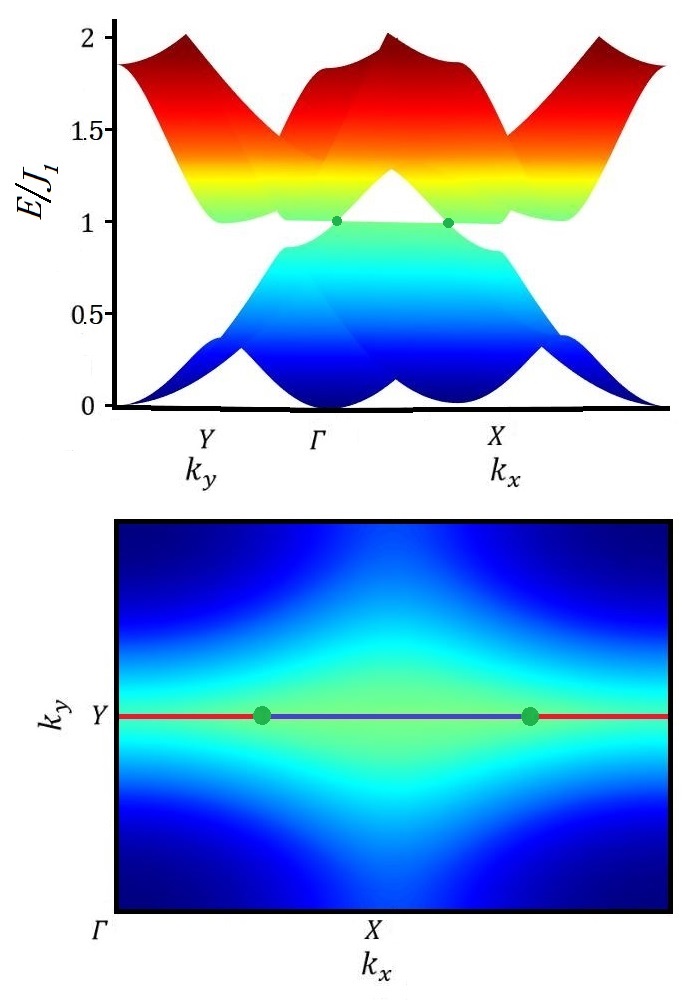}\label{fig4a}}
}
{
\subfigure[]{
\includegraphics[width=1.6 in]{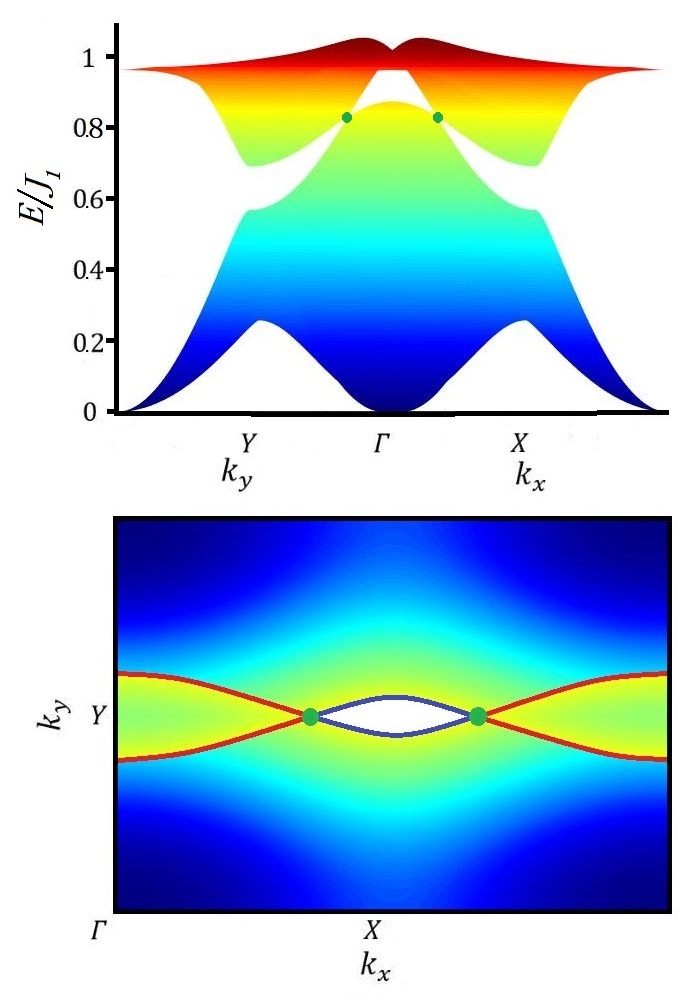}\label{fig4b}}
}

\caption{Comparison of the type-II Dirac magnons on the SHO lattice. (a) Constant energy cuts through the nodal energy with lines of bulk states between the nodes for the same parameters as Fig.~\ref{fig2b}; (b) Constant energy cuts with arcs open into bulk pockets for the bottom band and the second band. Parameter choices are $J_{1}$=1, $J_{2}$=0.5, $D_{1}$=0.15 and $D_{2}$=$D_{3}$=0.26.}
\label{Fig4}
\end{figure}

\subsection{Angular Momentum and Gyromagnetic Ratio}\label{subsec:AMGR}

\begin{figure*}[t]
\centering
{
\subfigure[]{
\includegraphics[width=1.6 in]{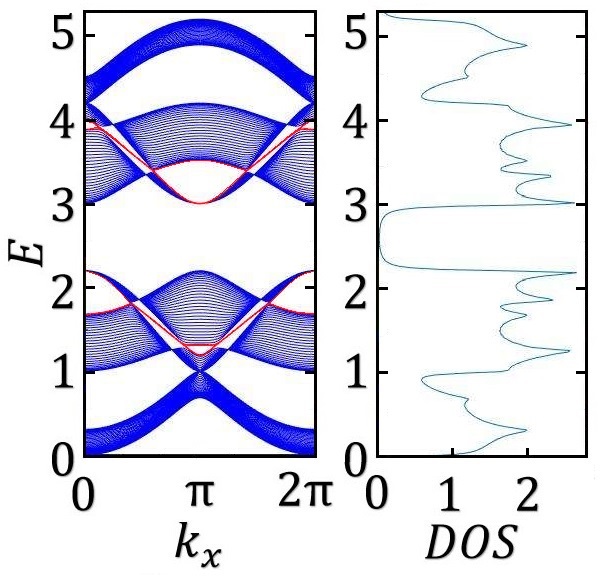}\label{fig5a}}
}
{
\subfigure[]{
\includegraphics[width=1.6 in]{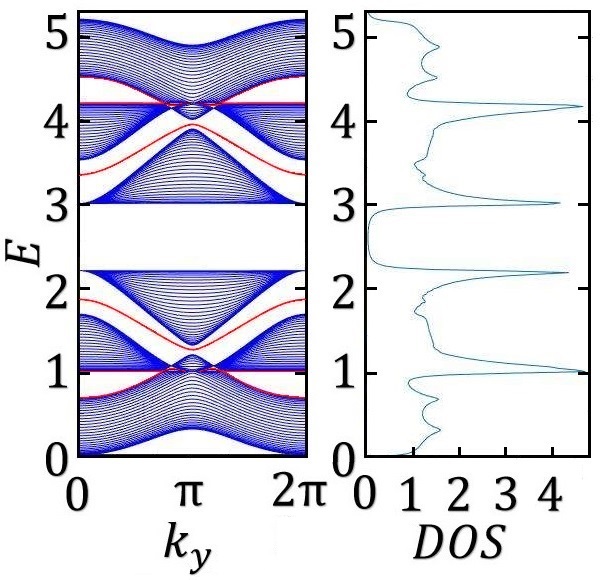}\label{fig5b}}
}
{
\subfigure[]{
\includegraphics[width=1.6 in]{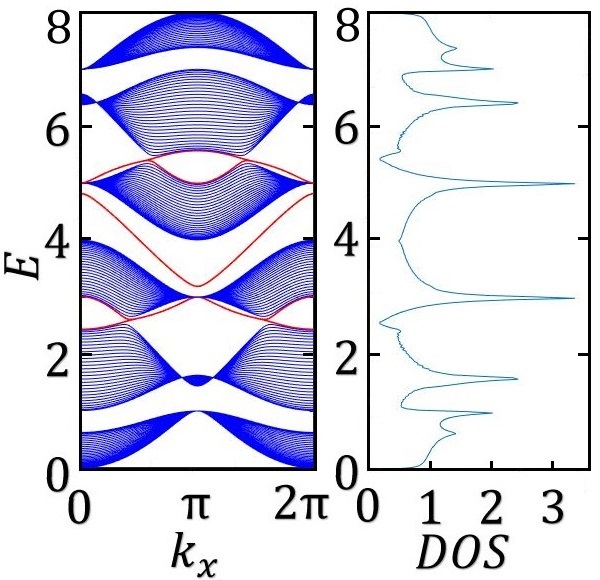}\label{fig5c}}
}
{
\subfigure[]{
\includegraphics[width=1.6 in]{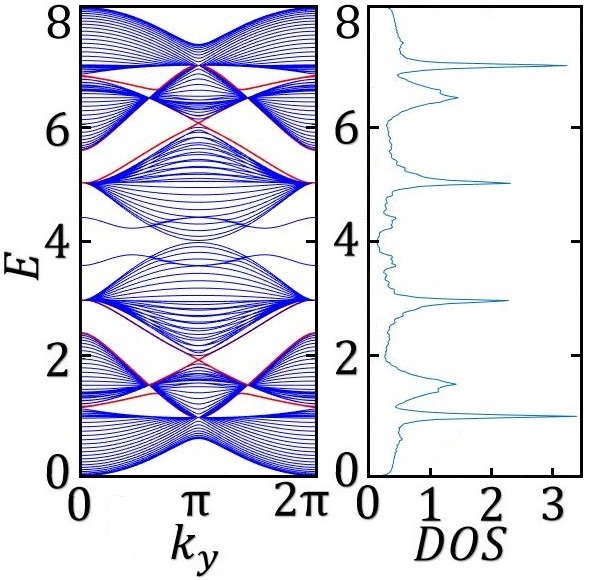}\label{fig5d}}
}
{
\subfigure[]{
\includegraphics[width=3.4 in]{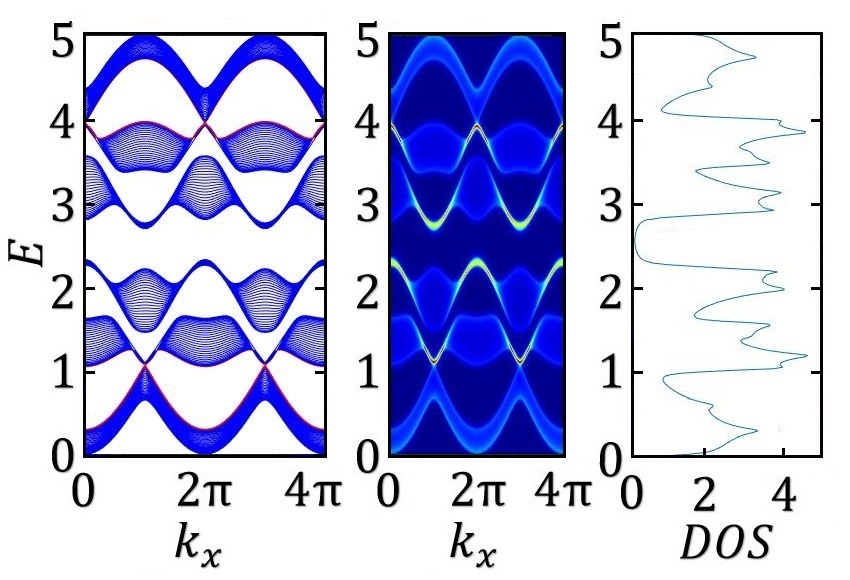}\label{fig5e}}
}
{
\subfigure[]{
\includegraphics[width=3.4 in]{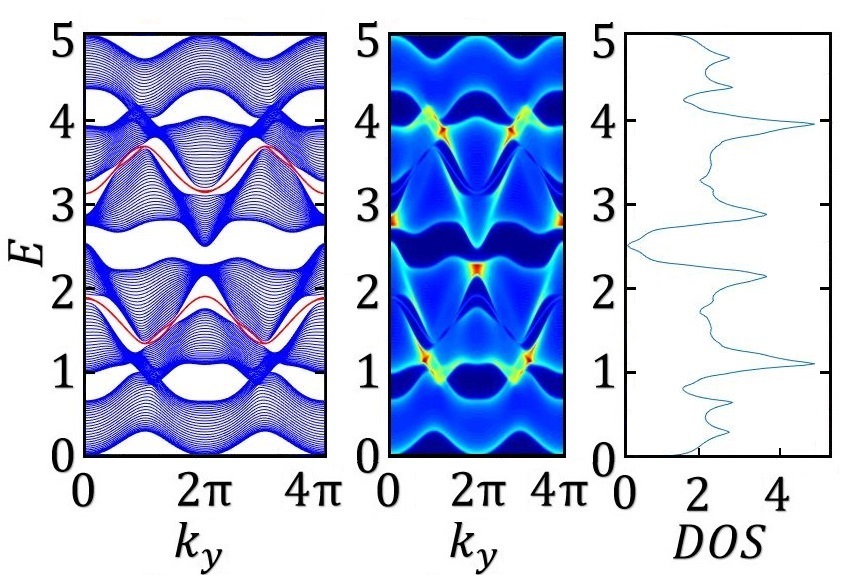}\label{fig5f}}
}

\caption{Chiral edge modes with energy set and magnon DOS on the SHO lattic. The parameters for the ribbon without DMIs: (a) The edge states and corresponding DOS along the $k_{x}$ direction with $J_{1}$=1, $J_{2}$=0.6 and $K$=$h$=0; (b) The edge states and corresponding DOS along the $k_{y}$ direction with $J_{1}$=1, $J_{2}$=0.6 and $K$=$h$=0; (c) The edge states and corresponding DOS along the $k_{x}$ direction with $J_{1}$=1, $J_{2}$=2 and $K$=$h$=0; (d) The edge states and corresponding DOS along the $k_{y}$ direction with $J_{1}$=1, $J_{2}$=2 and $K$=$h$=0; The parameters for the ribbon with $D_{1}$=$D_{2}$=$D_{3}$=0.1 where the Chern numbers are $\{ -1, -1, 2, 0, -1, 1\}$: (e) Topological edge modes with energy set and corresponding DOS along the $k_{x}$ direction with $J_{1}$=1, $J_{2}$=0.5 and $K$=$h$=0; (f)Topological edge modes with energy set and corresponding DOS along the $k_{y}$ direction with $J_{1}$=1, $J_{2}$=0.5 and $K$=$h$=0.}
\label{Fig5}
\end{figure*}

Considering the linear response theory, the transport coefficients for magnons consist of the deviations of a particle density operator and the current operators~\cite{PhysRevB84184406,PhysRevLett106197202}.
\begin{align}
&l_{edge} = \frac{2k_B}{4\pi^2 \hbar} 2\mathrm{Im} \sum_{n,\boldsymbol{k}} \big \langle \frac{\partial \psi_{n}} {\partial k_x} |Tc_1 (\rho(\varepsilon_{n \boldsymbol{k}})) -\frac{\rho(\varepsilon_{n \boldsymbol{k}}) \varepsilon_{n \boldsymbol{k}} }{k_B}|\frac{\partial\psi_{n}}{\partial k_y}\big \rangle,\non\\
&l_{self} = \frac{2k_B}{4\pi^2 \hbar} 2\mathrm{Im} \sum_{n,\boldsymbol{k}} \big \langle \frac{\partial\psi_{n}}{\partial k_x} |\frac{\rho(\varepsilon_{n \boldsymbol{k}})} {2k_B} (\varepsilon_{n \boldsymbol{k}} -H)|\frac{\partial\psi_{n}}{\partial k_y}\big \rangle.
\end{align}
where $c_1(x)$=$(1+x)\ln(1+x)$-$x\ln x$ is another weight function. The orbital motion of a magnon wave packet is defined as the total angular momentum without a mass term $\langle \boldsymbol{r} \times \boldsymbol{v} \rangle$ by summing the edge current and the self-rotation. Due to the lattice anisotropy, we introduce the effective mass tensor $\boldsymbol{m}^{\ast}$ describes how the mass of quasiparticles depend on the direction of motion.
\begin{align}
\frac{1} {\boldsymbol{m}^{\ast}} = \frac{1}{\hbar^2}\left[
\begin{array}{ccc}
\frac{\partial^2 E}{\partial {k_x^2}} & \frac{\partial^2 E}{\partial{k_x}\partial{k_y}} \\
\frac{\partial^2 E}{\partial{k_x}\partial{k_y}} & \frac{\partial^2 E}{\partial {k_y^2}} \\
\end{array}
\right],
\end{align}
Through the principal axis transformation, the anisotropic tensor can be diagonalized to the major $m_{maj}$ and minor $m_{min}$ axes. By analyzing the DOS near the relevant energy, the rank-2 tensor can be approximated to a scalar effective mass $m^{\ast}$ which is valid at low temperature due to the Bose-Einstein distribution.
\begin{align}
m^{\ast} = \sqrt{m_{maj}m_{min}},
\end{align}
Therefore, the total angular momentum per unit cell is given by
\begin{align}
L^{}_{tot} = m^{\ast}(l^{}_{edge}+l^{}_{self}),
\end{align}
where $L_{tot}$ represents the total angular momentum. Then the gyromagnetic ratio of magnons can be expressed as
\begin{align}
\gamma_m=\frac{\gamma_e L_{tot}}{\hbar \Delta m},
\end{align}
where the $\gamma_e$ is given by $2m_e/(ge)$, $g$ is the Lande factor, $e$ and $m_e$ are the charge and mass of the electron, respectively. As exotic quasiparticles in Boson systems, the gyromagnetic ratio response of topological magnons cannot be directly measured via traditional magnetometry in experiment. Thus, the differential response to temperature changes $\gamma_m^{\ast}$ provides indirect information about their gyromagnetic ratio.
\begin{align}
\gamma_m^{\ast}=(\frac{\partial L_{tot}/\partial T}{\partial \Delta m/\partial T})_{h},
\end{align}
By calculating the differential gyromagnetic ratio, we exhibit the EdH effect of topological magnons~\cite{PhysRevResearch.3.023248,PhysRevB.107.024408}.

\section{RESULTS}\label{sec:results}
\subsection{Type-II Dirac Magnons}\label{subsec:teb}

\begin{figure}[t]
\centering
{
\includegraphics[width=3.2 in]{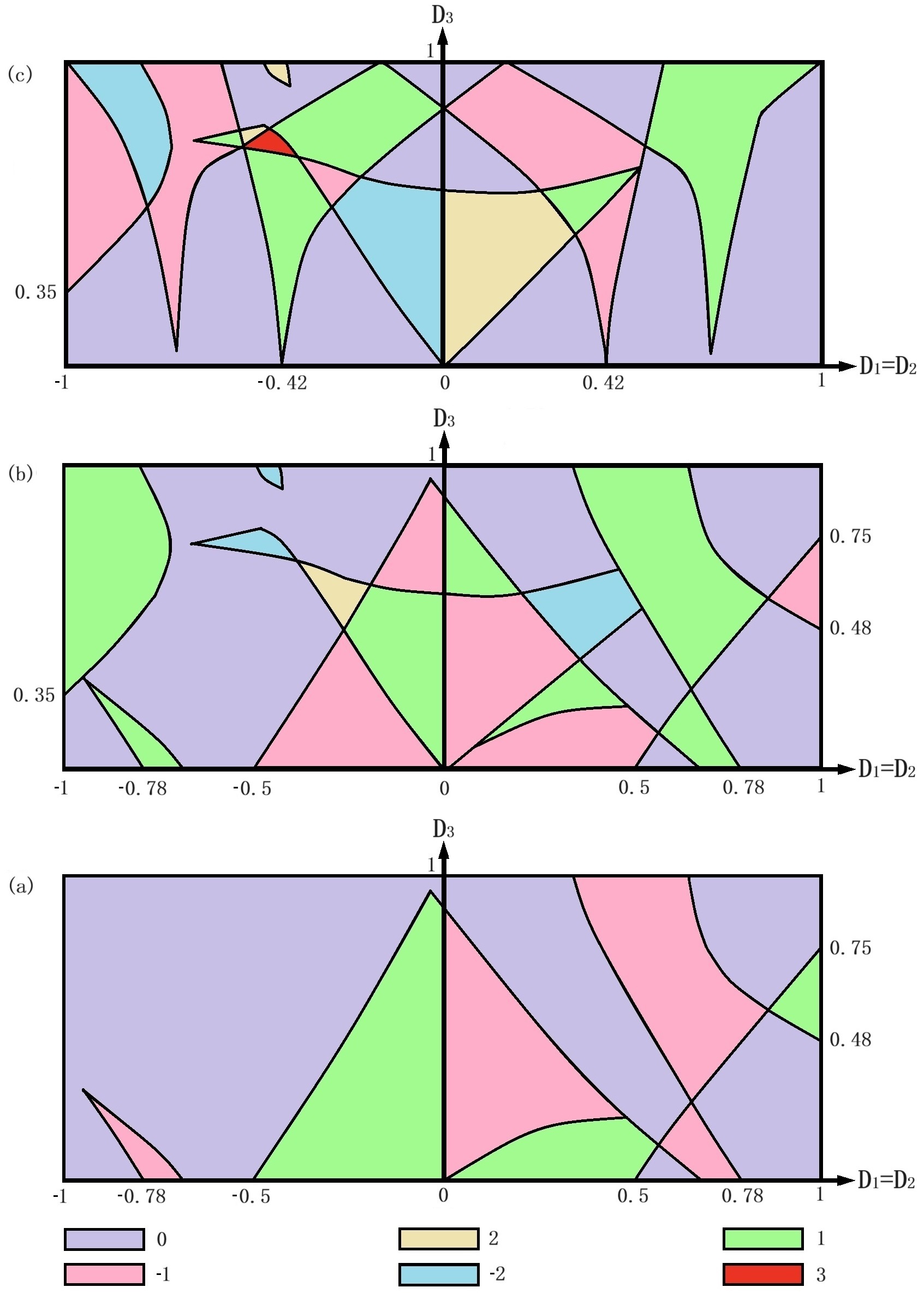}\label{fig11}
}
\caption{ Phase diagram of the Chern number for the first, second and third band (subfigure (a), (b) and (c)). We choose unified nearest-neighbor DMI as $D_1$=$D_2$ for $x$ axis and $D_3$ for $y$ axis. The fourth, fifth, sixth bands are mirror symmetric with the third, second and first band along the y-axis  respectively.}
\label{Fig11}
\end{figure}

Our study theoretically demonstrates the emergence of type-II tilted Dirac cones in the SHO lattice with anisotropic structure as shown in Fig.~\ref{Fig2}. The dispersion relation of these cones are tilted at an angle and no longer extend symmetrically in the momentum space. The type-II Dirac magnonic states are peculiar collective excitations owning anisotropic magnetization frequency branches, which result in the intersecting or merging of Dirac cones~\cite{RevModPhys.90.015001}. Comparing with the bands near the touching points in the type-I states, the connections between type-II Dirac magnons display linear or curved trajectories in nodal energy states. The arcs are preserved under small perturbations due to the saddle-shaped branches crossing the stable flat bands. The van Hove singularities(vHS) formed by the tilted Dirac cones act as the phase boundary in Fig.~\ref{Fig11}. Based on the calculation of the Wilson loop method, we obtain the topological Berry phase from a closed path surrounding the type-II Dirac crossing point. These points come in pairs of opposite chirality which must be brought together in momentum space and annihilated. Considering the anisotropy in two-dimensional BZ, the SHO lattice can be expected to have a variety of unique topological properties, including anisotropic magnon velocities and large non-reciprocal magnon transport.

In Fig.~\ref{fig4a} and Fig.~\ref{fig4b}, we adjust the exchange couplings inside and between hexagonal central cells to manipulate the band structure, which can drive the merging and generation of the type-II Dirac magnons. When the $J_2$ exchange couplings are smaller than 0.5 times of the $J_1$ exchange couplings, only the second and thrid bands (numbered from lower to higher) accompanied with their symmetrical counterparts are gapless. Within the range of 0.5 to 1, the increase of the ratio of $J_2$/$J_1$ weakens the potential between the bottem flat band and the saddle shaped band, leading to a shift of the Dirac points to the BZ center. The pair of Dirac cones annihilate at the saddle point when the ratio approaches 1, which eventually cause the topological transition and give the gap opening at the zone center. A zero $\mathbb{Z}_2$ index at the high symmetry point maps the disappearance of the nontrivial topology in the energy band. The sign of the potential changes from positive to negative indicating the vanishing of topologically protected edge modes between the type-II Dirac points, while the additional magnetization may be related to magnetically ordered states or magnon DOS. By manipulating the exchange couplings between and inside hexagonal central cells, these magnon modes provide potential means to develop novel and robust states in magnonics and spintronicss~\cite{PhysRevB.85.134411,science.1188260}.

\subsection{Topological Nontrivial Edge States}\label{subsec:dos}

We illustrate that the type-II Dirac states in the SHO lattice are topologically protected by $D_{2h}$ symmetry which induces chiral magnon edge modes and related non-trivial topological states. With increased magnetization, the pseudospins originated from orbital angular momenta of magnon modes exhibit an anisotropic set of edge states and unique magnon DOS. Preserving the lattice and the magnetic order, the chiral edge modes arise due to the breaking of inversion symmetry, which lead to the emergence of gapless states. We characterize the presence of edge modes via $\mathbb{Z}_2$ invariants resulting from the presence of inversion symmetry. Considering the interface between regions of topological and trivial domains, the nonlinear dispersions produce anisotropic magnetization frequency branches near the type-II Dirac points. The quantized nonzero Berry phases are exchanged at the tilted Dirac cones disclosing topological boundary states. During the jumpping of Berry phases, the domain boundaries can be considered as topological defects which disclose the topological transitions of the energy bands. The spin-momentum locking property stemmed from the coupling between the magnon bands offers a means of transporting and manipulating spin information~\cite{PhysRevB.48.11851,PhysRevB.90.024412}.

As shown in Fig.~\ref{fig5a} and Fig.~\ref{fig5b}, a new set of magnonic edge states emerge in lower energy bands via manoeuvring the Heisenberg exchange interactions. When the ratio of $J_2$/$J_1$ smaller than 0.75, the topologically protected edge states along the $k_x$ are localized to the bending of bulk bands near the boundary. Associated with the sharp peaks and edges of localized magnon DOS, the flat-saddle vHS states determine the local magnon transfer and convert the type-II Dirac states into topologically trivial states. The gapless states propagate along the emergence of exotic edge modes in Fig.~\ref{fig5c} and Fig.~\ref{fig5d},  while the $J_2$ reaches twice as much as the $J_1$. These modes can overlap partially in the gap between adjacent bands, which further enhances the topological anisotropy. We correlate them to the topological invariants providing a deeply understanding of the band phase diagram. Due to the increased robustness against disorders and perturbations, topologically protected chiral edge modes are immune to back-scattering.

\begin{figure}[t]
\centering
{
\subfigure[]{
\includegraphics[width=1.6 in]{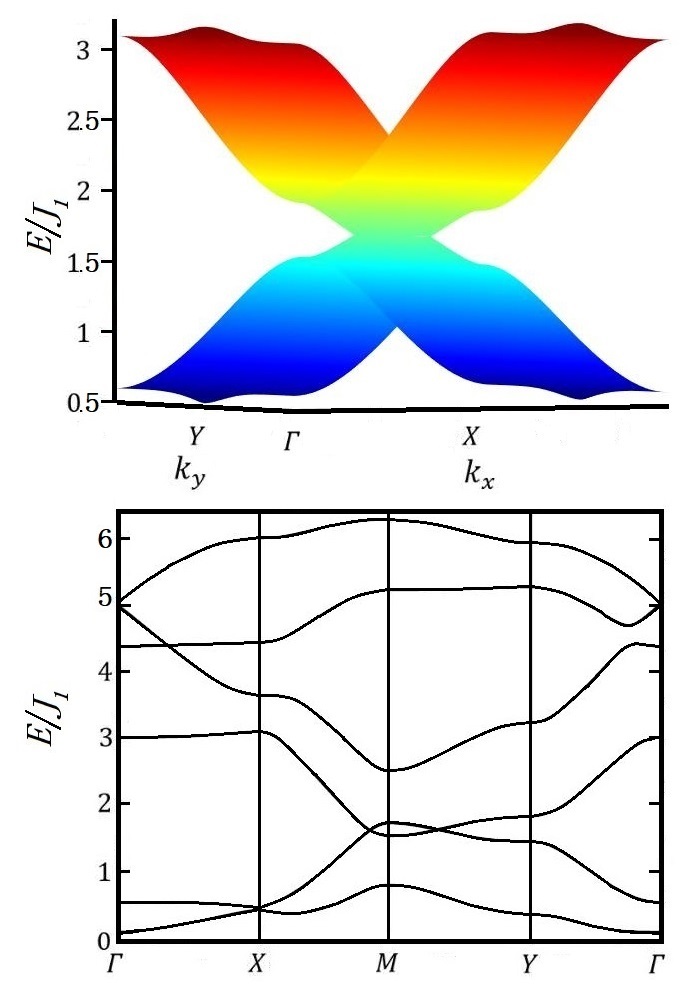}\label{fig6a}}
}
{
\subfigure[]{
\includegraphics[width=1.6 in]{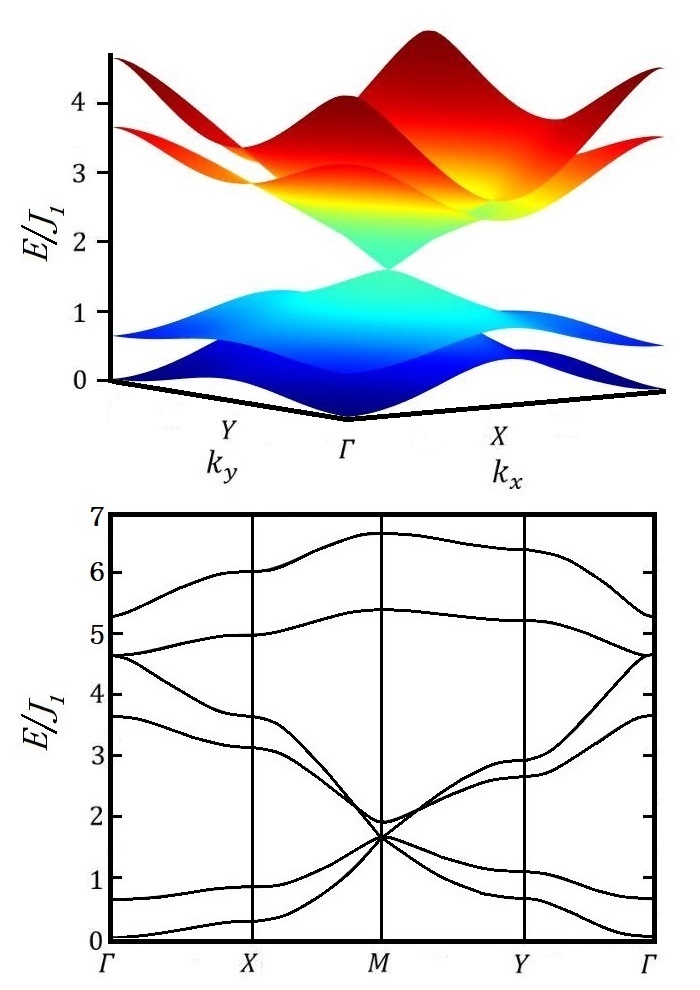}\label{fig6b}}
}
{
\subfigure[]{
\includegraphics[width=1.6 in]{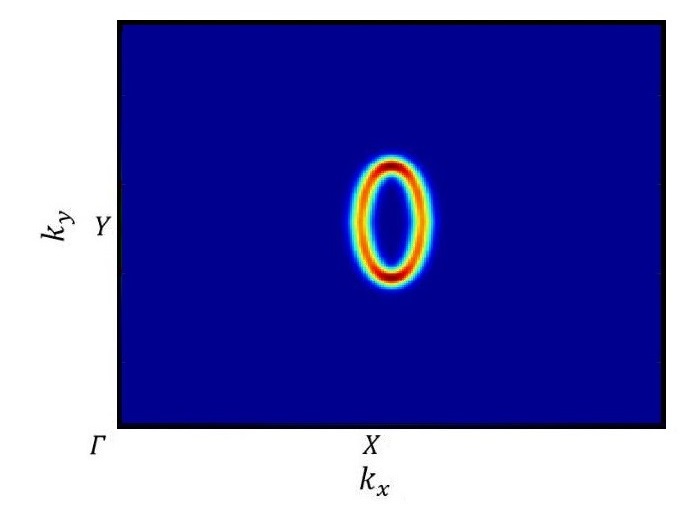}\label{fig6c}}
}
{
\subfigure[]{
\includegraphics[width=1.6 in]{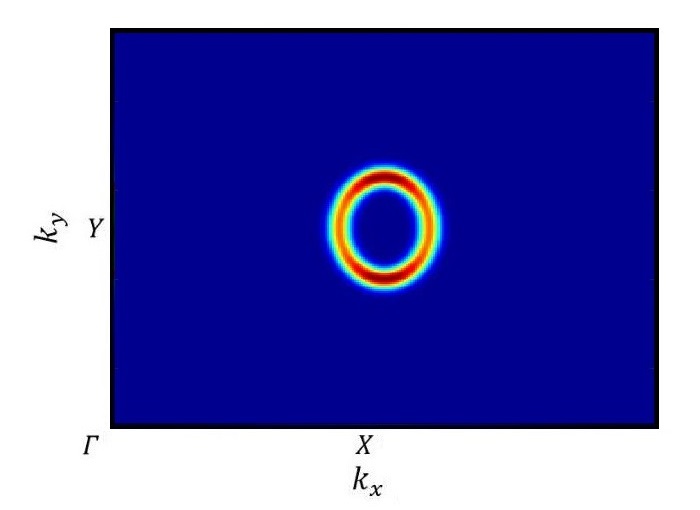}\label{fig6d}}
}

\caption{(a) Magnon band structures of Dirac Magnon Nodal Loops with $J_{1}$=0.5, $J_{2}$=1, $K$=0.2, $h$=0.25, $D_{1}$=0.4 and $D_{2}$=$D_{3}$=0.9. (b)Magnonic triply-degenerate points and the corresponding DNLs with $J_{1}$=0.5, $J_{2}$=1, $K$=0.2, $h$=0.25 and $D_{1}$=$D_{2}$=$D_{3}$=0.7. (c) The constant energy cut of the predicted experimental pattern at $E/J_{1}$=1.59 for the same parameters as Fig.~\ref{fig6a}. (d) The constant energy cuts through the corresponding DNLs $E/J_{1}$=2.15 for the inelastic neutron scattering for the same parameters as Fig.~\ref{fig6b}}
\label{Fig6}
\end{figure}

\subsection{Dirac Magnon Nodal Loops}\label{subsec:teb}

\begin{figure*}[t]
\centering
{
\includegraphics[width=7 in]{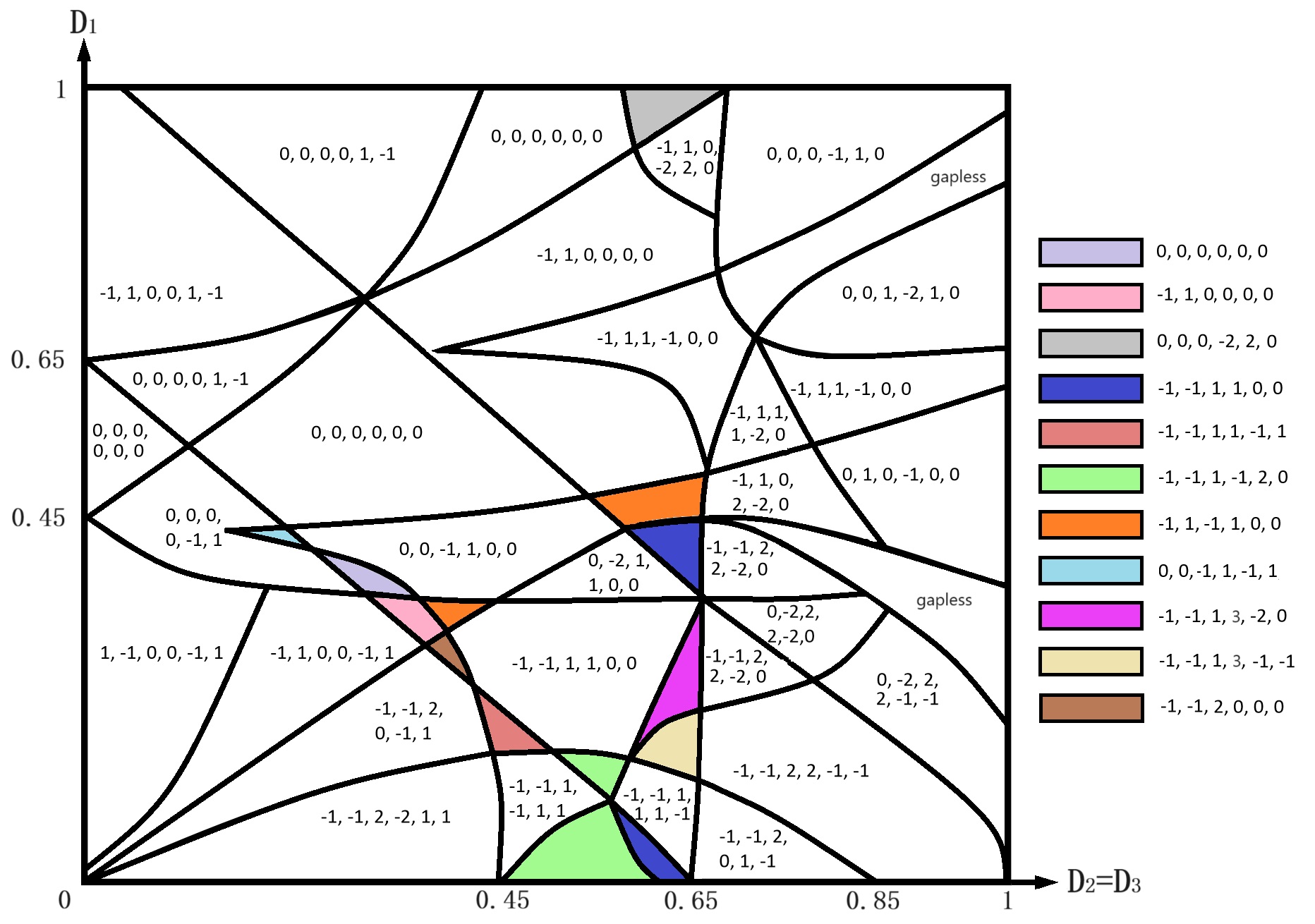}\label{fig7a}
}
\caption{Phase diagram of the SHO lattice. We choose unified octagon DMIs as $D_2$=$D_3$ for variable $x$ where $y$ variable is $D_1$. The index of the Chern number is for the first, second, third, fourth, fifth, sixth band (from lower to higher).}
\label{Fig7}
\end{figure*}
We find the direct magnonic analogues of Dirac nodal-line semimetals and Dirac triply-degenerate points (TPs) in 2D systems. In stark contrast to 3D systems, these magnonic analogues of DNLs are topologically protected by the $\mathbb{Z}_2$ invariant in the absence of any topological gap rather than transform into Weyl magnons. With the tendency to mix and overlap, the topological bands can shape 1D closed lines of Dirac nodes by applying appropriate DMIs. The TPs are formed by the crossing of three non-degenerate bands at the high symmetry $\pm M$ points. They have been proposed by theoretical studies on the magnons of composite lattice models~\cite{Owerre_2017,Owerre_2018} even if the DNLs and the TPs are elusive in electronic systems~\cite{ChinPhysLett2017}. We calculate the loop-integration of Berry phase for a closed path encircling the TPs, which is $\pm \pi$ while other paths without them is 0. While the existence of large DMIs leads to symmetry-breaking of the hexagonal sublattice, we have confirmed their topological protection through the Berry phase. As the magnetic excitations can be measured by inelastic neutron scattering, we present the constant-energy cuts of the predicted experimental pattern in Fig.~\ref{Fig6}. We have integrated over an energy window of $\pm$0.04$J_{1}S$.

\begin{figure*}[t]
\centering
{
\subfigure[]{
\includegraphics[width=2.2 in]{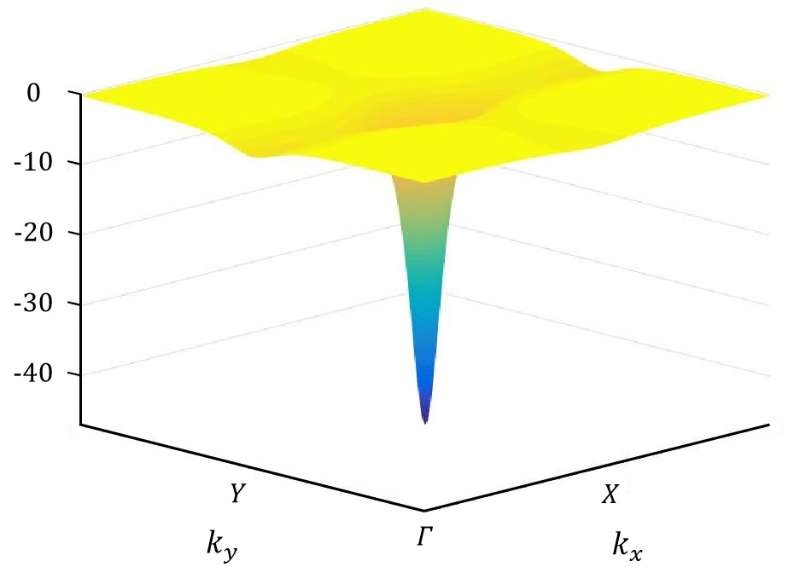}\label{fig8a}}
}
{
\subfigure[]{
\includegraphics[width=2.2 in]{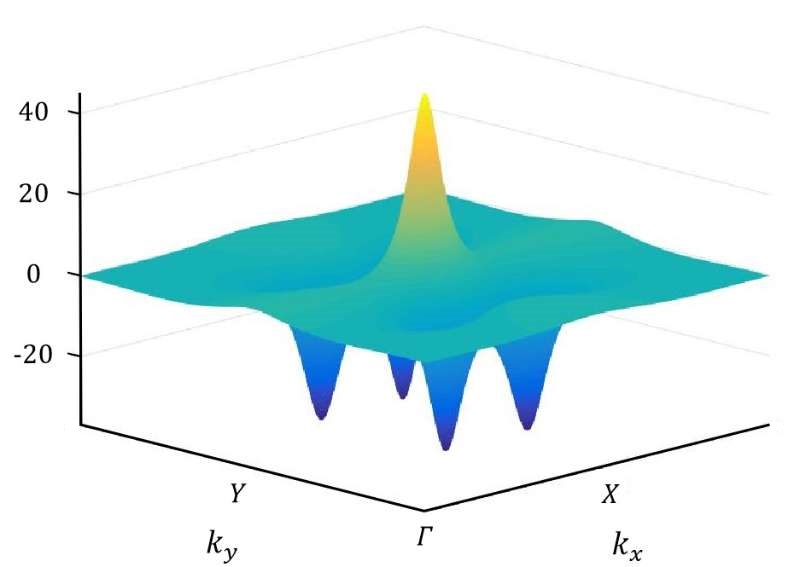}\label{fig8b}}
}
{
\subfigure[]{
\includegraphics[width=2.2 in]{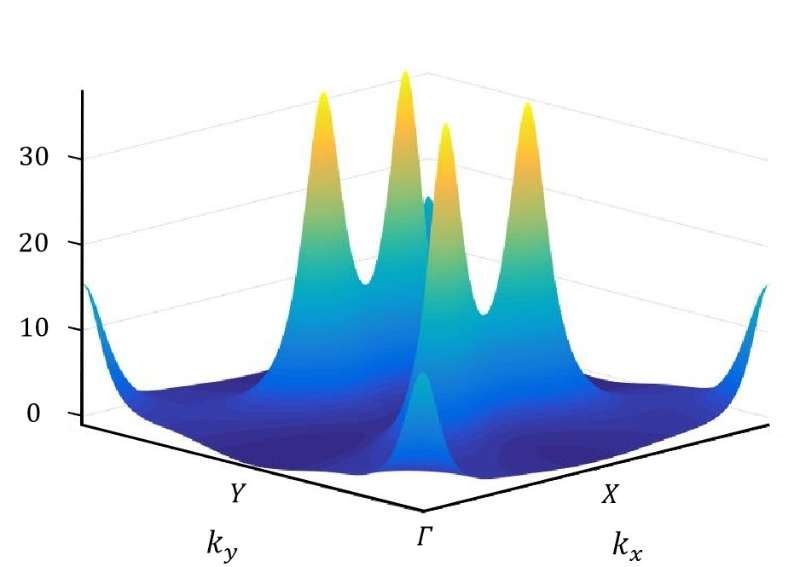}\label{fig8c}}
}
{
\subfigure[]{
\includegraphics[width=2.2 in]{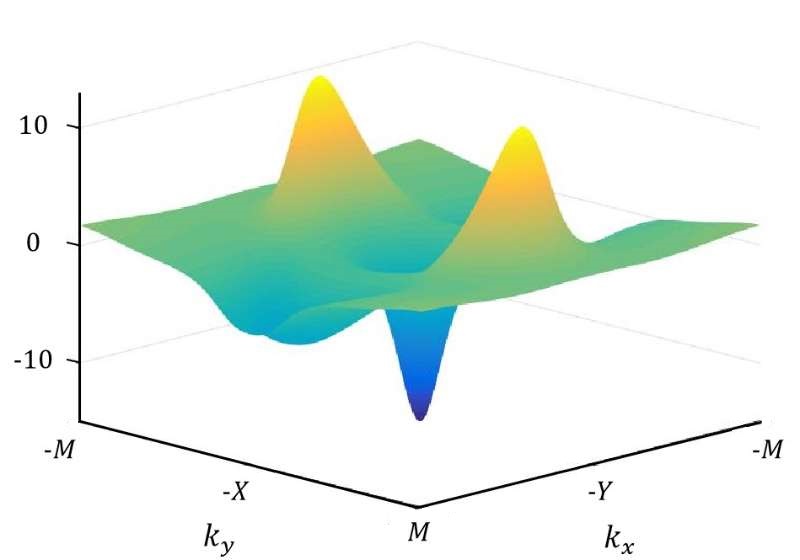}\label{fig8d}}
}
{
\subfigure[]{
\includegraphics[width=2.2 in]{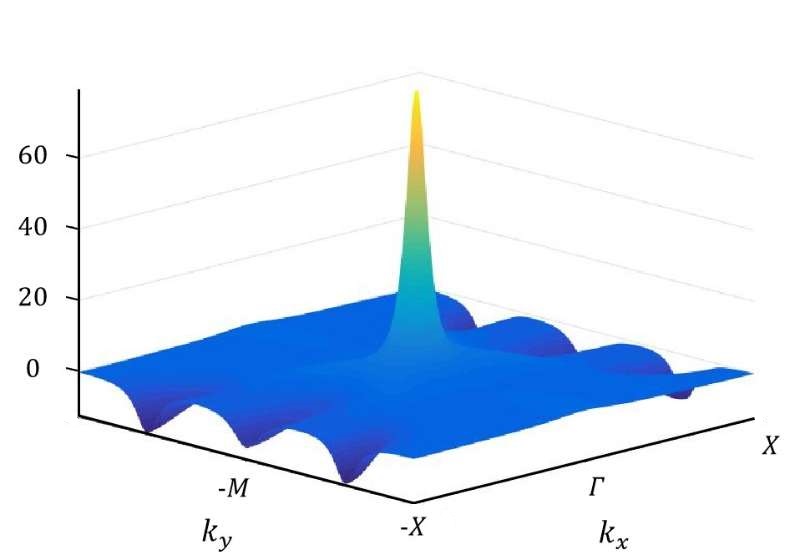}\label{fig8e}}
}
{
\subfigure[]{
\includegraphics[width=2.2 in]{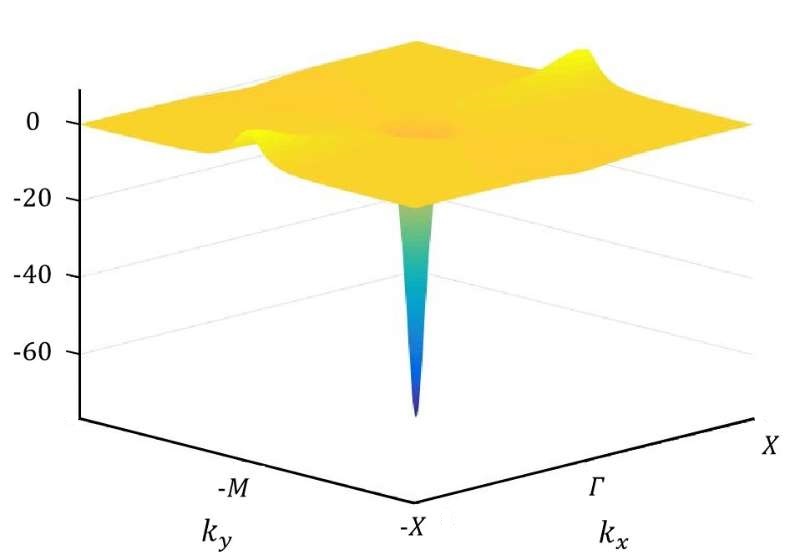}\label{fig8f}}
}
\caption{Berry curvature of magnon bands with $J_{1}$=1, $J_{2}$=0.5, $D_{1}$=0.2, $D_{2}$=0.4, $D_{3}$=0.2 and $K$=$h$=0.2. The (a), (b), (c), (d), (e), (f) figures correspond to the first, second, third, fourth, fifth, sixth band (from lower to higher), respectively. The Chern numbers are given by $\{ -1, -1, 2, 0, 0, 0\}$.}
\label{Fig8}
\end{figure*}

Moreover, the magnon bands feature a closed line of Dirac magnon nodes around the $\pm M$ points in Fig.~\ref{fig6a}. We confirm the topological protection from the four time-reversal-invariant momenta $\Gamma_{i}={\Gamma, X, Y, M}$  in the SHO lattice. Numerical calculation of the $\mathbb{Z}_2$ invariant is given by $(-1)^{\nu}$$=$$\Pi_{i=1}^{4}\Pi_{n=1}^{N}\xi_{n}(\Gamma_{i})$ where ${\nu}$=1 shows the nonzero $\mathbb{Z}_2$ invariant. The $\xi_{n}(\Gamma_{i})$ is the parity eigenvalue of the magnon bands forming the DNLs which are robust with two magnon bands overlap consisting the linear crossings. The presence of DNLs requires specific parameter range of DMIs, otherwise it shrinks to quadratic point node in the limit $\Delta D_2$$<$0.05 at $D_1$=$D_3$=0.4. As shown in Fig.~\ref{fig6b}, the collinear ferromagnet of the SHO lattice also exhibits magnon TPs at the $\pm M$ points, which are typically isolated points with a conical dispersion relation and additional linear term in the energy-momentum relation. The associated spin-wave velocities for the TPs at the $\pm M$ points along the two directions are $v_{kx}=\frac{1}{\hbar}\frac{\partial E}{\partial kx}$ and $v_{ky}=\frac{1}{\hbar}\frac{\partial E}{\partial ky}$. For the the first, second, third band (from lower to higher) the $v_{kx}=1.372J_{1}, 0.688J_{1}, 1.519J_{1}$ and the $v_{ky}=2.121J_{1}, 1.235J_{1}, 2.378J_{1}$ respectively. This unique dispersion exhibits a chirality imbalance. The corresponding gap states result in chiral magnon edge modes with finite thermal Hall effect due to broken time-reversal symmetry macroscopically. For comparison, we show the DNLs accompanied with the TPs for $D_1$=$D_2$=$D_3$=0.7, where the DNLs are robust in moderate DMIs regime. 

\subsection{Anomalous Thermal Hall Effect}\label{subsec:dos}

We achieve topologically distinct phases by modulating the nearest-neighbor DMIs and the next-nearest-neighbor DMI. The gapped magnon bands possess nonzero Berry curvature acting as an analogou of the effective magnetic field in momentum space~\cite{NPhys3347,FrontPhysChina}, which are convenient to manipulate both theoretically and experimentally. The rich topological phases coexist and compete with each other, which gives rise to a complex phase diagram of the Chern number. A thorough topological Chern phase diagram calculation is reported in Fig.~\ref{Fig7}. According to the bulk-boundary correspondence, there are loops of edge states in the band gap when the summation of Chern numbers are nonzero. We show the Berry curvature of magnon bands in Fig.~\ref{Fig8} with $J_{1}$=1, $J_{2}$=0.5, $D_{2}$=0.4, $D_{1}$=$D_{3}$=0.2 and $K$=$h$=0.2. The group velocity of the edge state shows a chirally moving magnon current localized at the edge, whose direction allows for the efficient and robust transport of magnons in the SHO lattice. Due to broken time-reversal symmetry macroscopically, we realize nontrivial topology by generating nonzero Berry curvature even when the integral of Chern number is zero. Accompanied with related transport properties, the DMI is key to investigate the angular momentum for topological edge current and self-rotation originating from the Berry curvature in magnon bands.

\begin{table}[htbp]
\centering
\caption{The Curie temperature $T_c$/$|J_1|$ for different parameters.}
\label{tab06}
\begin{tabular}{cccc ccccc}
\hline\hline\noalign{\smallskip}	
Lattice &  Parameters field & \textcolor[rgb]{1,0,0}{$T_c$/$|J_1|$}  \\
\noalign{\smallskip}\hline\noalign{\smallskip}
SHO & $J_1$=1, $J_2$=0.5, $D_1$=0.12, $D_2$=0.46, $D_3$=0.46, $K$=0.2 & 0.426 \\
SHO & $J_1$=1, $J_2$=0.5, $D_1$=0.2, $D_2$=0.63, $D_3$=0.63, $K$=0.2 & 0.368  \\
SHO & $J_1$=1, $J_2$=0.5, $D_1$=0.32, $D_2$=0.76, $D_3$=0.76, $K$=0.2 & 0.304  \\
SHO & $J_1$=1, $J_2$=0.5, $D_1$=0.72, $D_2$=0.4, $D_3$=0.4, $K$=0.2 & 0.337  \\
\noalign{\smallskip}\hline
\end{tabular}
\end{table}

\begin{figure}[t]
\centering
{
\includegraphics[width=3.6 in]{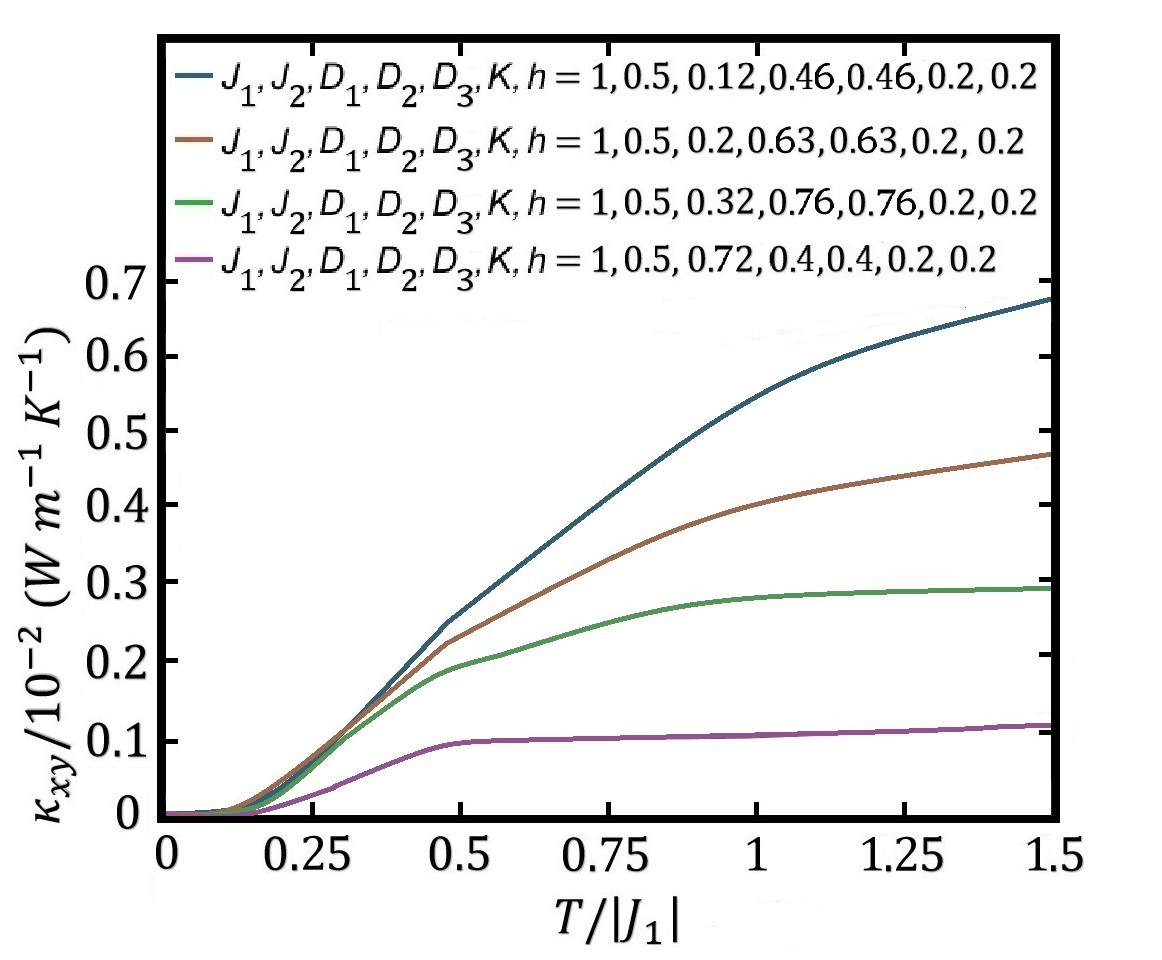}\label{fig9}
}
\caption{Low-temperature thermal Hall conductivity on the SHO lattice. In ferromagnetic coupling the DMIs, anisotropic term and Zeeman term are various with $J_1$=1, $J_{2}$=0.5 and $K$=$h$=0.2. The Chern numbers for the $D_1$=0.2, $D_2$=0.63, $D_3$=0.63 are given by $\{ -1, -1, 1, 3, -2, 0\}$. The Chern numbers for the $D_1$=0.12, $D_2$=0.46, $D_3$=0.46 are given by $\{ -1, -1, 1, -1, 1, 1\}$. The Chern numbers for the $D_1$=0.32, $D_2$=0.76, $D_3$=0.76 are given by $\{ 0, -2, 2, 2, -2, 0\}$ and the Chern numbers for the $D_1$=0.72, $D_2$=0.4, $D_3$=0.4 are given by $\{ -1, 1, 0, 0, 0, 0\}$}
\label{Fig9}
\end{figure}

As shown in Fig.~\ref{Fig9}, the topological thermal Hall conductivity undergoes a platfrom as a consequence of the type-II Dirac state, which vanishes at zero temperature. When the energy gap between the lowest magnon excitation and the rest of the magnon bands is large enough, the low-energy excitations can be considered as an independent sector, and the thermal Hall conductivity is determined solely by the properties of this sector, such as the Berry curvature. When the energy gap is small or comparable to the thermal energy, thermal excitations across the gap become important, and the thermal Hall effect is dominated by these excitations instead of the low-energy magnons. Therefore, the observation of a large thermal Hall effect at low temperatures can be used as a probe of the magnon band structure and the presence of a significant energy gap. The strength of the thermal Hall conductivity is proportional to the Berry curvature and is thus determined by the topological properties of the system.

\subsection{Einstein-de Haas Effect}\label{subsec:EdH}

\begin{figure}[t]
\centering
{
\subfigure[]{
\includegraphics[width=3.6 in]{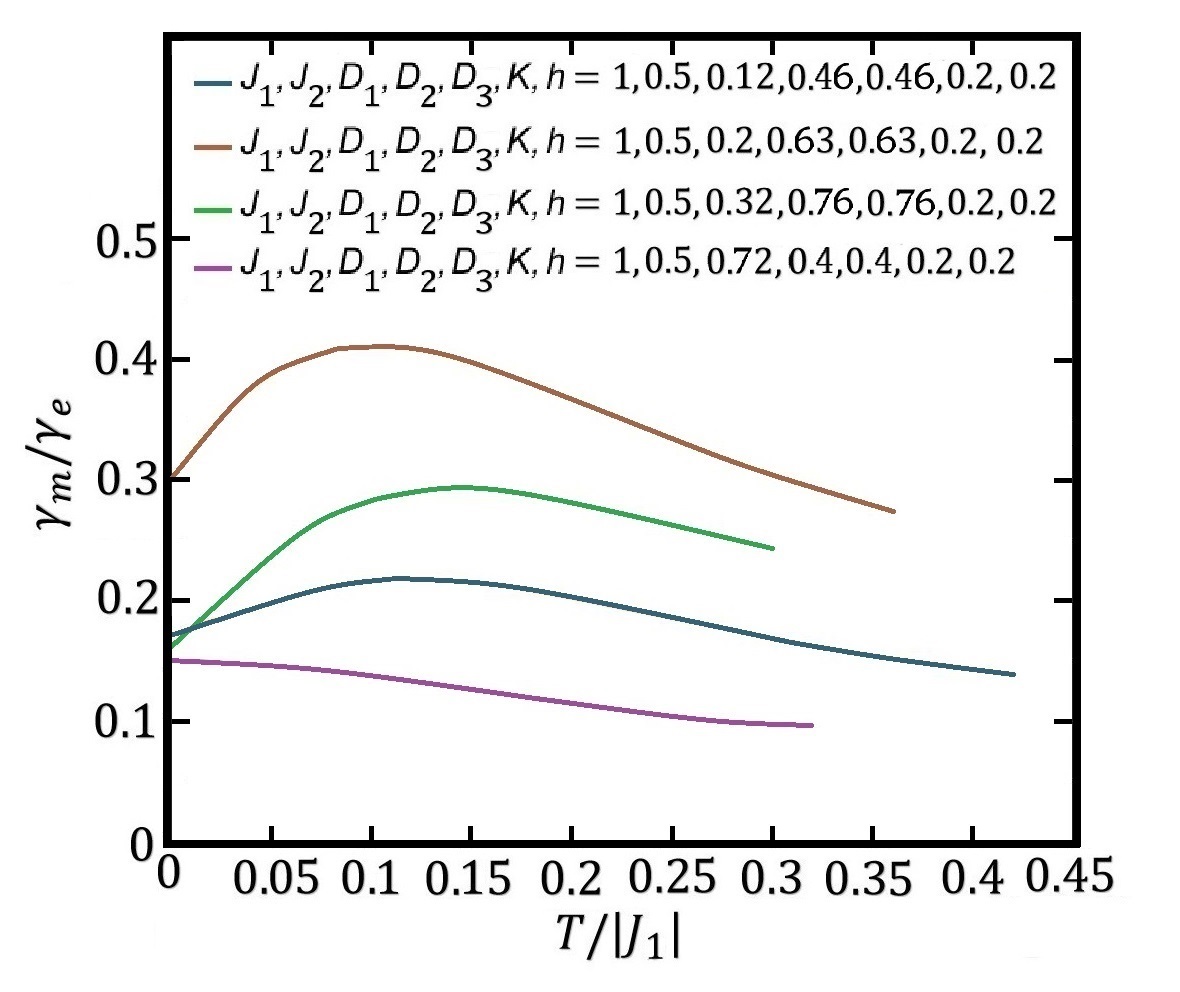}\label{fig10a}}
}
{
\subfigure[]{
\includegraphics[width=3.6 in]{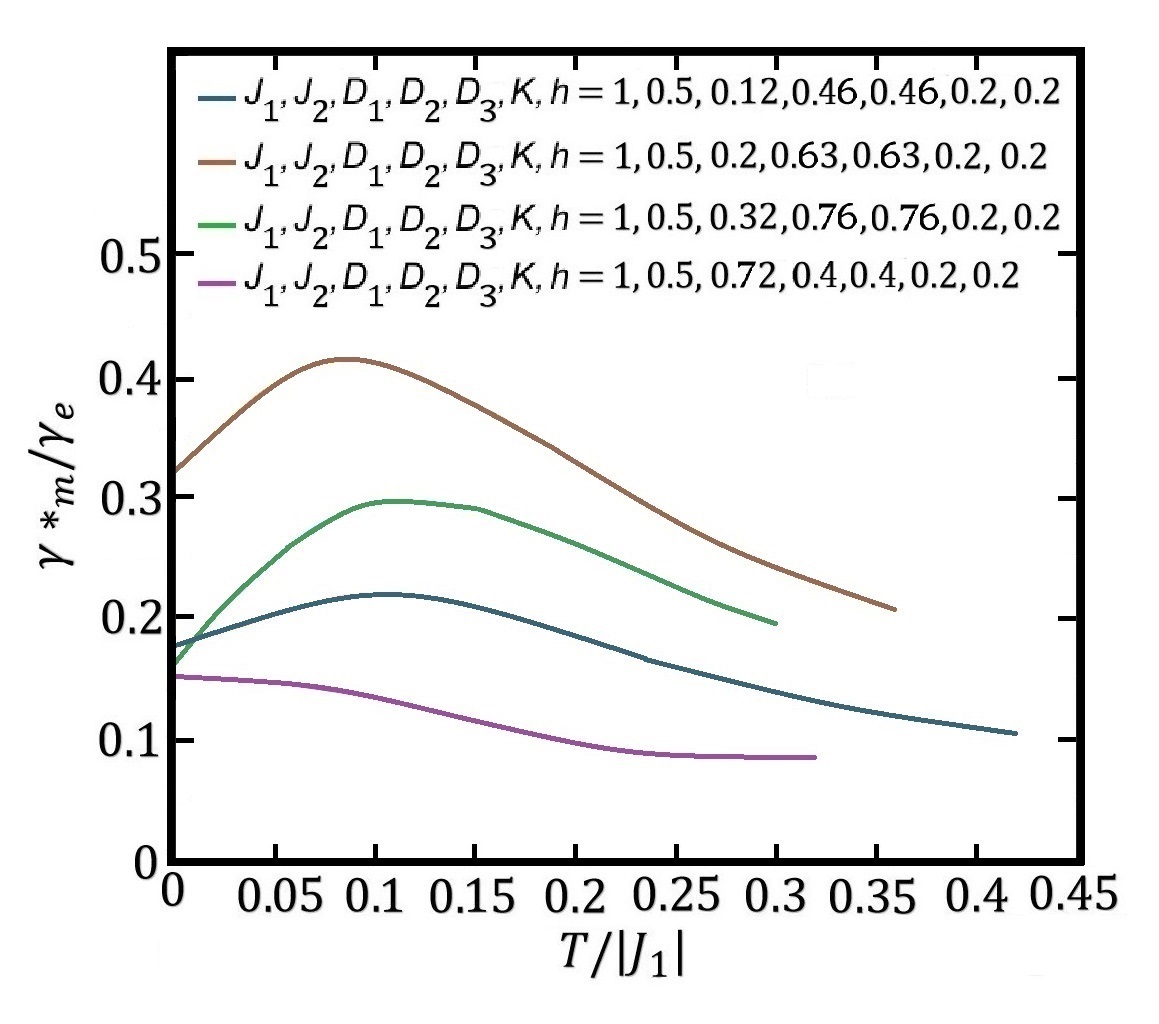}\label{fig10b}}
}
\caption{Comparison of the Einstein de-Haas effect response on the SHO lattice with different parameters. (a)The topological gyromagnetic ratio with $J_1$=1, $J_{2}$=0.5. (b)The differential gyromagnetic ratio variation with temperature is shown. Parameter choices are the same as before.}
\label{Fig10}
\end{figure}

According to the linear response theory, the magnon wave packets undergo two types of orbital motions and the total angular momentum is defined as the summation of these two types of rotational motions~\cite{PhysRevB84184406,PhysRevLett106197202}. We define the gyromagnetic ratio as the angular momentum divided by the magnetic moment of magnons, which is related to the magnetization change of the system. Each magnon mode can be excited or annihilated and has its own gyromagnetic response. Our results refer that the total gyromagnetic contribution increases significantly at first and reaches a peak value at about $T$=0.10$|J_1|$. To further analyze the physical content of the EdH effect, we compare the gyromagnetic ratio of the SHO lattice system with different DMIs parametric system. We show the results of our calculation in Fig.~\ref{Fig10}. The Einstein-de Haas effect is a special example of the more general phenomenon of angular momentum conservation~\cite{NewJPhys.20.103018}. In this case, the transfer of angular momentum can be achieved through a variety of means, including magnetization, torque, and rotational motion~\cite{PhysRevB.103.L100409,PhysRevLett.125.117209}. It has been used to measure the magnetization of materials, determine spin-decoherence times in ferromagnetic materials, and study the effects of magnetic fields on superconductors~\cite{PhysRevLett.129.167202}. Other applications of this effect include the development of gyroscopes and magnetometers.

The $\gamma_m/\gamma_e$ shown in Fig.~\ref{fig10a} represents the temperature variation of the topological gyromagnetic ratio compared to the electronic value. As the $\gamma_e^{\ast}$ is equal to $\gamma_e$ for electrons, the $\gamma_m^{\ast}/\gamma_e^{\ast}$ can be simplified as $\gamma_m^{\ast}/\gamma_e$. Hence, the differential gyromagnetic ratio is renormalized from the $\gamma_m$ response. Considering the differential gyromagnetic ratio response, the magnon system also has a peak value before descending as seen in Fig.~\ref{fig10b}. Thus, there is an optimal temperature of the differential gyromagnetic ratio at which the magnon insulator will have the strongest response. From an experimental point of view, there is an optimal temperature zone in which our theory can be tested well. The values of optimal temperature for various SHO systems are all from $T=0.05|J_1|$ to $T=0.20|J_1|$.

\section{CONCLUSIONS}\label{sec:conclu}

In summary, we have proposed the concept of the type-II Dirac magnons along with methods to realize topological changes of energy bands without requiring the DMIs. Then, we have revealed unique topological features on the SHO lattice which resembles a distorted kagome lattice with flat-saddle bands. The anisotropic magnon band structure has unique saddle-shaped bands and flat bands preserving the inversion symmetry and reflection symmetry. The intersections between the saddle-shaped bands and flat bands display magnonic arcs tracing the nodal energy of the type-II Dirac points. With nonlinear relationship between the spin magnetization energy and the incident angle, the type-II Dirac magnonic states possess mass terms with opposite sign in the trivial and topological domains. Furthermore, we have explored moderate nearest-neighbor Heisenberg exchange to lengthen the magnonic arcs and strengthen the energy difference, thereby enhancing topologically non-trivial magnon modes. The near-critical tilted Dirac cones merge together at the saddle point exhibiting a topological transition, where the magnon DOS shows a sharp peak. The discrete jump of the Berry phase describing the topological protection of the degenerate magnon bands characterize various low-energy features including nontrivial topology changes.

Similar to the type-II Dirac fermions, the magnon excitations exhibit magnonic analogues of the zone-center vHs states~\cite{PhysRevB.98.094419,PhysRevB.89.134409}. With adjusting the $J_2$/$J_1$, the Dirac point and its time-reversal partner approach each other and eventually merge together at the $\Gamma$-point. This merge of the Dirac points results in a topological transition, making the system topologically trivial. The type-II Dirac magnons are gapped by the DMIs acting as an effective magnetic field to induce the nonzero Berry curvature. Our findings show that the topological magnon bands are stable under the DMIs as long as the ground state symmetry is preserved. As rarely found in simple 2D systems, the DNLs semimetals are currently attracting widespread interest in condensed matter physics~\cite{PhysRevB.95.014418}. The Dirac magnon nodal-line loops can be controlled while the frequency of the magnon modes can then be adjusted accordingly. Our work opens an avenue for another new topological distinction of magnonic analogue of DNLs semimetals which are robust for the DMIs within a certain parameter range. We calculate the expected finite frequency intensity of the DNLs based on the bulk sensitivity of the inelastic neutron scattering methods~\cite{PhysRevB.78.052507}. The inelastic neutron-scattering studies can be expected to distinguish the Dirac node-line semimetals with the inclusion of DMIs in magnonic systems~\cite{PhysRevB.97.134411}. We elucidate the topological nature of these new states and suggest potential applications in spintronic and magnonic devices~\cite{PhysRevLett.115.106603,PhysRevB.91.125413}.

Our study provides a experimental realization of the EdH effect, which is a macroscopic mechanical manifestation caused by the angular momentum conservation. The Stewart's apparatus is an original device designed to directly measure the EdH effect, which consists of a metal cylinder suspended by a wire. By passing an electric current through the coil, the magnetic field created produces a torque on the cylinder, causing it to rotate. This angular momentum can then be measured in terms of the angle of rotation. Recently, an experimental setup has been proposed to detect the EdH effect for the collective excitation modes of topological magnons~\cite{PhysRevResearch.3.023248}. A disk shaped magnetic sample is connected to a suspension wire to provide the torque and measure the topological differential gyromagnetic ratio response. Refering to the unusual propagation through the energy bands, the EdH effect support the applications of spintronics where the frequency of the magnon mode can be used to operate the flow of spin-polarized currents~\cite{PhysicsReports2022,PhysRevLett.108.246402}. Besides, it is very interesting to study the topological properties of magnons on the SHO lattice which can be realized in magnetic materials. These magnon states may also be observed in topology circuit which should be achieved by depositing magnetic atoms on a metallic substrate using the STM technique. Meanwhile, the tuning of exchange couplings proposed on the SHO lattice can be realized in Rydberg atom by trapping and manipulating the magnetic atoms in an optical lattice. Moreover, the EdH effect can produce mechanical effect which has potential applications in quantum informatics and topological magnon spintronics~\cite{PhysicsReports2022}.


\begin{acknowledgments}
We would like to thank Jun Li, Xin-Wei Jia and Zenan Liu for helpful discussions. This project is supported by NKRDPC-2022YFA1402802, NKRDPC-2018YFA0306001, NSFC-92165204, NSFC-11974432, Leading Talent Program of Guangdong Special Projects (201626003), and Shenzhen Institute for Quantum Science and Engineering (Grant No. SIQSE202102).
\end{acknowledgments}

\bibliography{cite}
\end{document}